\documentclass[a4paper]{article}
\usepackage{arxiv}

\usepackage[utf8]{inputenc} 
\usepackage[T1]{fontenc}    
\usepackage{hyperref}       
\usepackage{url}            
\usepackage{booktabs}       
\usepackage{amsfonts}       
\usepackage{nicefrac}       
\usepackage{microtype}      
\usepackage{graphicx}
\usepackage{doi}
\usepackage{Commands}

\title{On a kinetic opinion formation model for pre-election polling}

\date{}

\author{\href{https://orcid.org/0000-0002-3601-2869}{\includegraphics[scale=0.06]{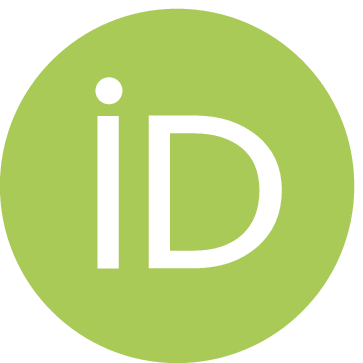}\hspace{1mm}Bertram D\"uring}\\
	Mathematics Institute\\
	University of Warwick\\
	Zeeman Building \\
	Coventry, CV4 7AL \\
	United Kingdom \\
	\texttt{bertram.during@warwick.ac.uk} 
	\And
\href{https://orcid.org/0000-0002-6365-2554}{\includegraphics[scale=0.06]{orcid}\hspace{1mm}Oliver Wright} \\
	Mathematics Institute\\
	University of Warwick\\
	Zeeman Building \\
	Coventry, CV4 7AL \\
	United Kingdom \\
	\texttt{oliver.l.j.wright@warwick.ac.uk}
}

\renewcommand{\headeright}{}
\renewcommand{\undertitle}{}
\renewcommand{\shorttitle}{Kinetic opinion formation model for pre-election polling}

\hypersetup{
pdftitle={On a kinetic opinion formation model for pre-election polling},
pdfsubject={physics.soc-ph,nlin.AO}
pdfauthor={Bertram D\"uring, Oliver Wright},
pdfkeywords={opinion formation, demographics, pre-election polling, Boltzmann equation, Fokker–Planck equation},
}

\begin{document}
\maketitle

\begin{abstract}
\noindent Motivated by recent successes in model-based
pre-election polling, we propose a kinetic model for opinion formation which includes voter demographics and socio-economic factors like age, sex, ethnicity, education level, income and other measurable factors like behaviour in previous elections or referenda as a key driver in the opinion formation dynamics. The model is based on \textit{Toscani}'s kinetic opinion formation model \cite{bib:Toscani:kineticmodel} and the leader-follower model of \textit{D\"uring et al.} \cite{bib:During:strongleaders}, and leads to a system of coupled Boltzmann-type equations and associated, approximate Fokker-Planck-type systems. Numerical examples using data from general elections in the United Kingdom show the effect different demographics have on the opinion formation process and the outcome of elections.
\end{abstract}

\keywords{opinion formation, demographics, pre-election polling, Boltzmann equation, Fokker–Planck equation.}

\section{Introduction}
   
\noindent Different kinetic models to describe opinion formation in a society consisting of a large number of interacting individuals have been proposed and analysed in the literature, see e.g.\  \cite{bib:Toscani:kineticmodel,bib:During:strongleaders,bertotti2008discrete,motsch2014heterophilious,zbMATH06350891,bib:During:inhomogeneous,albi2016opinion,bib:ToscaniPareschi:ims,toscani2018opinion,albi2015optimal,albi2016recent,zbMATH06665924,zbMATH06553232,zbMATH06684855,zbMATH06684856,zbMATH06712347,zbMATH06735322,zbMATH06875734,zbMATH06892489,zbMATH06977672,zbMATH07205672,zbMATH06944946,albi2019boltzmann,zbMATH07146312,burger2021network}.
Building on mathematical approaches from statistical mechanics they frequently lead to generalizations of the classical Boltzmann equation for gas dynamics. Instead of binary collisions, one considers the process of compromise between two individuals. Similar models are used in the modelling of wealth and income distributions which show Pareto tails, cf.\ \cite{bib:During:economykineticmodel} and the references therein. A key application of such kinetic opinion formation models is political opinion formation in elections or referenda. When aiming to push the boundaries of where such methods can be applied to analyse or forecast election results (see also \cite{galam2008sociophysics}) it is useful to look at recent developments in pre-election polling.

Pre-election polling which aims to forecast election outcomes based on surveys of political opinion used to rely on relatively small samples. But a small number of observations at national level make it difficult to forecast outcomes at sub-national (constituency) level, leading to frequent (real or perceived) failures of pre-election polls \cite{bib:FT}, 
e.g.\ in 2015 and 2017 UK general elections.
Even if a pre-election poll correctly forecasts the share of a party at national level, who wins the election in a first-past-the-post electoral system depends on the outcome on the local (constituency) level, i.e.\ who wins each of the many seats in parliament. In recent years, so-called {\em multilevel regression and post-stratification} (MRP) methods have shown to be able to form high-quality sub-national estimates of political opinions by uncovering dependencies of voter behaviour on demographic, socio-economic and other measurable factors \cite{park2004bayesian}. MRP methods are much better at capturing the complex nuances of opinions depending on local variations in historic voter behaviour and voter demographics. To this end they use a larger poll size so that sufficient samples from each constituency are gathered, and at the same time record demographic and socio-economic characteristics like age, sex, ethnicity, education level, income and information on previous votes in elections or referenda from the participants. Then a regression is performed to determine how likely it is that an individual with a certain combination of demographic and socio-economic characteristics votes for a particular party. In a post-stratification step, census and other publicly available data is used to determine the composition of the voter population in each constituency with regard to these demographic and socio-economic characteristics and then forecast the outcome of the vote in each of the constituencies.

In this paper, motivated by the development and successes of MRP methods for pre-election polling, we propose and analyse a kinetic model for opinion formation which takes into account demographic and socio-economic factors like age, sex, ethnicity, education level, income and other measurable factors like behaviour in previous elections or referenda. 
The basic underlying kinetic model opinion formation is the
leader-follower model from \cite{bib:During:strongleaders} which in turn is based on
the opinion formation model by Toscani \cite{bib:Toscani:kineticmodel}. Mathematically, it is related to works in the
kinetic theory of granular gases \cite{cercignani2013mathematical}.
In the kinetic opinion formation models the evolution of the distribution of opinion is
described by means of microscopic interactions among individuals in a
society. These microscopic interactions are reminiscent of collisions
in gas dynamics. Opinion is represented as a continuous variable $w\in
\mathcal{I}$ with $\mathcal{I} = [-1, 1]$ representing the left–right political
spectrum, and $\pm 1$ represent extreme opinions. In each interaction
two main drivers affect an individual's opinion. The first one is a
compromise process \cite{hegselmann2002opinion,deffuant2002can,weidlich2006sociodynamics},
in which individuals tend to reach a
compromise after exchange of opinions. The second one is
self-thinking, where individuals change their opinion in a diffusive
way, possibly influenced by exogenous information sources such as the
media.
In this setting, the time evolution of the distribution of opinion
among individuals is governed by (systems of)
Boltzmann-type equations. In a suitable
quasi-invariant limit, partial differential equations of Fokker–Planck type
can be derived for the distribution of opinion. 

The paper is organized as follows.
In the next section we propose a generalised kinetic opinion formation
model which takes into account demographic and socio-economic factors in the compromise process of the
kinetic opinion formation model, and present the resulting system of
Boltzmann-type equations and its associated quasi-invariant limit
Fokker-Planck system. In Section~\ref{sec:Deriv} we propose an approach to use real-world pre-election polling and census data to inform the choice of compromise functions in our model. We put this to test in two
real-world examples from the 2019 United Kingdom general elections,
for the East of England region and Nottinghamshire which forms part of
the so-called ``red wall'' of constituencies in northern England which
was widely perceived to have played a key role in the outcome of the 2019 elections.
We present results from direct Monte Carlo simulations of the
microscopic model and compare it with finite element based simulations
of the Fokker-Planck system. Section~\ref{sec:Conclusion} concludes.

\section{Derivation of the Model} \label{sec:Deriv}
    \subsection{Interactions} \label{sec:Deriv:Interactions}


        We propose a model based on \textit{Toscani}'s opinion formation model \cite{bib:Toscani:kineticmodel} and the leader-follower model of \textit{D\"uring et al.} \cite{bib:During:strongleaders}, where we look at the distribution of the individuals' opinion, $w$, in an interval $\mathcal{I}=[-1,1]$. We split our population into $N$ follower species and a leader species, which we label as species $1,2,...,N$ and species $L$ and associate with a density function, $f_i(w, t)$ and $f_L(w,t)$, respectively, depending on the opinions, $w\in \I$, and time, $t\ge0$. Each follower species will be associated with a single demographic. Each demographic can include numerous demographic and socio-economic factors like age, sex, ethnicity, education level, income and other measurable factors like behaviour in previous elections or referenda. To limit the number of species to a practical value, we do not include all possible demographic factors, but consider only the factors that are assumed to be most important in driving opinion formation \cite{LAUDERDALE2020399}.
    
     The dynamics of our system are defined by looking at binary interactions (or collisions) between two individuals from our total population. We define the interactions in this model, in three possible cases: (a) the interactions where a member of the leader species interacts with another member of the leader species (an $L$ and $L$ interaction); (b) a member of the leader species interacts with a member of the follower species $i$ for some $i=1,...,N$ (an $L$ and $i$ interaction); and (c) a member of a follower species interacts with another member of a follower species (an $i$ and $j$ interaction for some $i,j = 1,...,N$). We assume that the leader species is formed by strong opinion leaders, as in \cite{bib:During:strongleaders}, that is assertive individuals able to influence followers but withstand being influenced themselves. To that end, we have that a leader species will have their opinion unchanged by an interaction with an individual from a follower species, but interact normally with a another leader. Thus we look at the following cases for the interactions with $w_L^*, v_L^*$ and $w_L, v_L$ as the post- and pre-interaction leader opinions, respectively $w^*, v^*$ and $w, v$ as the post- and pre-interaction opinions of the follower species, respectively:
    
    \begin{enumerate}[(a)]
        \item The $L$ and $L$ case:
            \begin{equation} \label{int:LL}   
                \begin{split}
                    w_L^* &= w_L+ \gamma_L P_{LL}(w_L,v_L) (v_L-w_L) + D_L(w_L) \eta_L, \\ 
                    v_L^* &= v_L+ \gamma_L P_{LL}(v_L,w) (w_L-v_L) + D_L(v_L) \eta_L;
                \end{split}
            \end{equation}
        \item The $i$ and $L$ case, $i = 1,2,...,N$:
            \begin{equation} \label{int:cL}   
                \begin{split}
                    w^* &= w + \gamma_i P_{iL}(w, v_L) (v_L-w) + D(w) \eta_i, \\ 
                    v_L^* &= v_L;
                \end{split}
            \end{equation}
        \item The $i$ and $j$ case, $i,j = 1,2,...,N$:
            \begin{equation} \label{int:cc} 
                \begin{split}
                w^* &= w + \gamma_i P_{ij}(w,v) (v-w) + D(w) \eta_i, \\ 
                v^* &= v + \gamma_j P_{ji}(v,w) (w-v) + D(v) \eta_j.
                \end{split}
            \end{equation}
    \end{enumerate}
    
    In the interactions above, $\gamma_\nu \in \real$, for $\nu = 1, 2,... N, L$, denotes compromise parameters which govern the ``speed'' of attraction of two different opinions, the functions $P_{ij}(\cdot, \cdot)$ and $D(\cdot)$ model the local relevance of compromise and self-thinking for a given opinion, respectively, and $\eta_\nu$ are the ``self-thinking" random variables, which model how the opinion of an individual changes independently from other individuals. Here, we have made the implicit assumption that the total number of followers and leaders in a species -- and by extension the mass of the density functions -- remain constant over time. We only mention here that when using this to model a phenomenon with a timescale of years, demographic changes (migration to other species, birth/death) could also be incorporated by adding gain and loss terms similarly to \cite{bib:During:strongleaders}.

    To ensure that $w^*$ and $v^*$ remain in $\I$, we choose $P_{ij}$ and $D$ functions accordingly. Our $P_{ij}$ functions are chosen in such a way that they remain in $[0,1]$. In the prevalent literature, this has been taken to be a localisation function, such as a characteristic function or a hyberbolic tangent, \cite{bib:Toscani:kineticmodel, bib:During:strongleaders, bib:During:inhomogeneous}, here we propose a more realistic function based on real-world polling data in Section~\ref{sec:Deriv}.\ref{sec:Deriv:PFunction} below. Then, $P_{ij}(w,v)(v-w)\in [-2,2]$, hence we choose $\gamma_\nu \in (0,{1}/{2})$, so that $-1<\gamma_\nu P_{ij}(w,v)(v-w) <1$. The function $D$ is chosen in such a way as to ensure that the diffusion term is sufficiently small near the boundaries which guarantees that the post-interaction opinions remain in $\I$. Thus we require the function $D$ to remain in $[0,1]$ and $D(-1) = D'(-1) = D(1) = D'(1) = 0$. We also want $D$ to be symmetric in this case, although this is not required. A usual choice for this function is: $$D(x) = (1-x^2)^\alpha,$$ for some $\alpha>0$. We choose $D_L$ similarly and for both $\alpha = 2$, unless otherwise stated.

    Finally, we now discuss the random variables $\eta_\nu$. Let us define a set of probability measures in the following way, for some $\delta>0$: 
    $$\mc{M}_{2+\delta} = \left\{ \mathbb{P} : \mathbb{P} \text{ is a probability measure and, } \mean{|\eta|^\alpha} =  \int_\I |\eta|^\alpha \ \od \mathbb{P}(\eta)<\infty, \forall \ \alpha \le 2+\delta  \right\}.$$
        
    We choose a probability density $\Theta \in \mc{M}_{2+\delta}$ and we assume that this density is obtained from a random variable $Y$ with the properties,  $\mean{Y} = \int_\I Y \ \od \Theta (Y) = 0$, $\mean{Y^2} = \int_\I Y^2 \ \od \Theta (Y) = 1$, then we choose $\eta_\nu$ with the property that $\mean{|\eta_\nu|^p} = \mean{|Y\sigma_\nu|^p} $ for all $ \nu = 1, 2,..., N,L$ and $p \in [0,2+\delta]$, with $0<\sigma_\nu < 1$. We assume that the opinion of an individual is more likely to be swayed by an interaction with another individual rather than their exposure to outside factors, thus $\sigma_\nu \ll 1$.


    \subsection{Boltzmann-type and Fokker-Planck equations} \label{sec:Deriv:BoltFokker}

    
        From the microscopic interactions in the previous subsection $N+1$ Boltzmann-type equations can be derived using standard methods of kinetic theory \cite{bib:During:strongleaders, bib:Cercignani:RGD}. These come in two forms, $N$ equations that describe the behaviour of the follower species,
    
    \begin{equation} \label{eq:Boltz}
        \npderiv{t}f_i(w, t) = \frac{1}{2\tau_{i L}} Q(f_i, f_L)(w,t) + \sum_{j = 1}^N \frac{1}{2\tau_{i j}} Q(f_i, f_j)(w,t),\quad \text{for } i = 1,2...,N, 
    \end{equation}
    and the equation that describes the behaviour of leader species,

    \begin{equation} \label{eq:BoltzL}
        \npderiv{t} f_L(w,t) = \frac{1}{2\tau_{L}} Q(f_L, f_L)(w,t),
    \end{equation}
    with the interaction operators $Q(f_\mu, f_\nu)(w,t)$ between species $\mu, \nu =1, ... , N, L$ in \eqref{eq:Boltz} and \eqref{eq:BoltzL}, expressed in weak form as
    \begin{align*}
        \int_{\I} Q(f_\mu, f_\nu)(w,t) \phi(w) \ \od w = \Big \langle \int_{\I^2} \big[ \phi(w^*) +& \phi(v^*) - \phi(w) - \phi(v) \big]  f_{\mu}(w,t) f_\nu(v,t) \ \od w \od v \Big \rangle,
    \end{align*}
    where $\phi$ are test functions from $C_c(\I;\real)$, that are continuous and compactly supported functions on $\I$. In \eqref{eq:Boltz} and \eqref{eq:BoltzL}, we use $\tau_{ij}$, $\tau_{iL}$ and $\tau_L$ to denote the relaxation times of the Boltzmann-type equations which allow to control the interaction frequencies. 

    We use similar methods to those described in \cite{bib:During:strongleaders} to derive a set of Fokker-Planck equations for this model by taking a quasi-invariant limit of \eqref{eq:Boltz}, \eqref{eq:BoltzL}. It is well-known that these quasi-invariant limits are a good approximation for the stationary profiles of the Boltzmann-type equation \cite{bib:Toscani:kineticmodel}. To study these large time stationary distributions, therefore, we use the transformation $s=\gamma t$, for some $\gamma \ll 1$, writing the large time transformed density functions $g_\mu (w,s) = f_\mu(w,t)$ for every $\mu$, using the scaling $\gamma_\nu = \alpha_{\nu} \gamma$ and $\sigma_\nu = \beta_{\nu} \sigma$ for all $\nu = 1,...,N,L$, and restricting our test functions to twice differentiable functions, all of whose derivatives are $\delta$-H\"older continuous for some $\delta>0$ with compact support, that is $\phi \in C^{2,\delta}_c(\I,\real)$. In the limit $\gamma,\sigma\to 0$ while keeping $\lambda=\sigma^2/\gamma$ fixed, this yields the following nonlinear, nonlocal Fokker-Planck equations for the follower species,
    \begin{equation}
        \begin{split}
            \npderiv{s} ( g_i(w,s)) =&  \npderiv{w}\Bigg[ \bigg(\frac{\alpha_i}{2\tau_{iL}} \mc{K}_{L,iL}(w,s) + \sum_{j=1}^N \frac{\alpha_j}{2\tau_{ij}} \mc{K}_{j,ij}(w, s)\bigg)  g_i(w,s) \Bigg] \\
            & +\sum_{j=1}^N \frac{\alpha_{j}}{2\tau_{j}} \npderiv{w}\bigg( \mc{K}_{i,ij}(w,s) g_j(w,s)  \bigg) \\
            & + \left( \frac{\lambda_i M_L(s)}{4\tau_{iL}} + \sum_{j=1}^N \frac{\lambda_j M_j(s)}{4\tau_{ij}} \right) \nsecpderiv{w}\bigg( D^2(w) g_i(w,s) \bigg) \\
            & + \sum_{j=1}^N \frac{\lambda_j M_i(s)}{4\tau_{ij}} \nsecpderiv{w} \bigg( D^2(w)  g_j(w,s) \bigg), \quad \text{for } i = 1,2,...,N,
        \end{split} \label{eq:Fokker}
    \end{equation}
    and, similarly, the Fokker-Planck equation for the leader species $L$:
    \begin{equation}
        \begin{split}
            \npderiv{s}  ( g_L(w,s)) =&  \frac{\alpha_{L}}{\tau_{L}} \npderiv{w}\Bigg(  \mc{K}_{L,LL}(w,s)   g_L(w,s) \Bigg)  \\
            & +\frac{\lambda_L M_L(s)}{2\tau_{L}}  \nsecpderiv{w}\bigg( D_L^2(w) g_L(w,s) \bigg),
        \end{split} \label{eq:FokkerL}
    \end{equation}
    where $\lambda_\nu = {\beta^2_\nu\lambda}$ for all $\nu = 1,..,N,L$ and with $\xi, \mu, \nu = 1,2,...,N,L$,
    \begin{equation*}
        \mc{K}_{\xi,\mu \nu}(w,t) := \int_\I P_{\mu \nu}(w,v)(v-w) g_\xi(v,t) \ \od v, \ M_{\xi}(s) := \int_\I g_\xi(w,s) \ \od w.
    \end{equation*}
    We now show that the density masses remain constant for all $s>0$ in the Fokker-Planck model, that is $M_\zeta(s) = M_\zeta(0)$ for all $s>0$ and $\zeta = 1,2, ... ,N$. Consider the following by substituting the weak form of \eqref{eq:Fokker} into $\nderiv{s} \int_{\I}g_\zeta(w,s)\phi(w) \ \od w$,
    \begin{align}
        \nderiv{s}\bigg(M_\zeta&(s)\bigg) = \nderiv{s} \int_\I g_\zeta(w,s) \phi(w)\ \od w  \\
        =& \int_\I  \bigg(\frac{1}{2\tau_{\zeta L}} \mc{K}_{L,L\zeta}(w,s) + \sum_{j=1}^N \frac{1}{2\tau_{\zeta j}} \mc{K}_{j, \zeta j}(w, s)\bigg)  g_\zeta(w,s) \phi '(w) \ \od w \nonumber \\ 
        & +\int_\I \sum_{j=1}^N \frac{\alpha_{\zeta j}}{2\tau_{\zeta j}} \mc{K}_{\zeta,\zeta j}(w,s) g_j(w,s)  \phi'(w) \ \od w \nonumber \\ 
        & +\int_\I \left( \frac{\lambda_\zeta M_L(s)}{4\tau_{\zeta L}} + \sum_{j=1}^N \frac{\lambda_\zeta M_j(s)}{4\tau_{\zeta j}} \right) D^2(w) g_\zeta(w,s) \phi''(w) \ \od w \nonumber \\
        & + \int_\I \sum_{j=1}^N \frac{\beta^2_{\zeta j} \lambda_\zeta M_\zeta(s)}{4\tau_{\zeta j}}  D^2(w)  g_j(w,s) \phi''(w) \ \od w. \label{eq:FokkerBounds}
    \end{align}
    
    A choice for the test function, consistent with the conditions we have placed on it, is $\phi(w) \equiv 1$, which leads directly to $\nderiv{s}(M_\zeta(s)) = \nderiv{s} \int_\I g_\zeta(w,s) \ \od w = 0$ and thus to $M_\zeta(s) = M_\zeta(0)$ for all $s>0$. Analogously, we obtain $M_L(s) = M_L(0)$ for all $s>0$.
    
    From this point, it will be prudent to rename the function $g_\nu(w,s)$ back to $f_\nu(w,t)$ for all $\nu = 1,2,...,N,L$ for ease of presentation.

    \subsection{Choice of compromise function from demographic voter intention data} \label{sec:Deriv:PFunction}

    
    In \cite{bib:During:inhomogeneous, bib:During:strongleaders, bib:Toscani:kineticmodel}, the compromise function $P$ has been chosen as a mere localisation function, usually a hyperbolic tangent function or a characteristic function for some radius $r$. Although this approach has been used to great effect when considering two party systems \cite{bib:During:inhomogeneous}, we propose here a more data-driven approach to include demographic and other factors into the compromise functions, which is more suited to a multi-party political system. To achieve this, we look to voter intention polls made before the election, thus, as briefly explained in Section~\ref{sec:Deriv}.\ref{sec:Deriv:Interactions}, we propose a new form of $P$ which is based on that polling data. Each follower species will be associated with one combination of $n$ ``characteristics", $c_{i1}, c_{i2}, ... , c_{in}$, each of which is chosen from a potential set of mutually exclusive groups, that is all individuals in a species belong to one, and only one, group, and we assign to all our species one of these groups. Using the UK General Elections as an example, we could choose the characteristics of age and home-ownership. These have mutually exclusive groups that could be associated with them, that is under 24 years-old, and over 24 years old for age, and either a home-owner or not for home-ownership. We then use four species to which we assign the groups, under 24 home-owners, over 24 home-owners, under 24 non-home-owners and over 24 non-home-owners. To this end, we make the assumption that $P_{ij}(w,v)$ can be decomposed into a localisation function $P_{loc}$, which depends only on the opinions, and characteristic comparison functions $P_{char}$, which depend on the opinions and, in addition the data from pre-election polls \cite{bib:YouGov:voterintention} for each of the characteristics of the species $i$ and $j$.

    Let $w,v$ be the opinions of two individuals in species $i$ and $j$, respectively, with associated characteristic groups $c_{ik}, c_{jk}$ for all $k=1,...,n$, then we use the following compromise drift function decomposition:
        \begin{equation}
            P_{ij}(w,v) = P_{loc}(w, v) \prod_{k=1}^n P_{char}(w, v;  c_{ik}, c_{jk}).
        \end{equation}

    We use a characteristic function as our localisation function in this paper,
    \begin{equation*}
        P_{loc}(w,v) = \mathbbm{1}_{\{|w-v| < r\}},
    \end{equation*}
    for some $r>0$. We use only $P_{loc}$ for the leader species that is $P_{iL} (w,v) = P_{LL}(w,v) = P_{loc}(w,v) $.

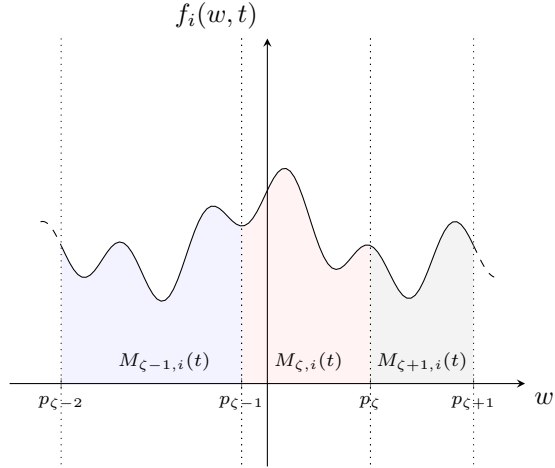
\begin{figure}[h]
    \begin{center}
        \begin{tikzpicture}
    	    \begin{axis}[
        		axis x line=center,
    	    	axis y line=center,
    		    xlabel={$w$},
    			ylabel={$f_i(w,t)$},
    			xlabel style={below right},
    			ylabel style={above left},
    			xticklabels ={,,},
    			yticklabels = {,,},
    			xtick={-1, -0.125, 0.5, 1},ytick={\empty},
    			xmin=-1.25,
    			xmax=1.25,
    			ymin=-0.3,
    			ymax=1.25]
    
    			\addplot [name path=f,mark=none,domain=-1:1,samples=100,yticklabels=\empty] {0.1*cos(deg(x*3.141593))+ 0.1*cos(deg(2*x*3.141593)) + 0.1*sin(deg(5*x*3.141593))+0.5};
                \addplot [dashed, mark=none,domain=1:1.1,samples=100,yticklabels=\empty] {0.1*cos(deg(x*3.141593))+ 0.1*cos(deg(2*x*3.141593)) + 0.1*sin(deg(5*x*3.141593))+0.5};
                \addplot [dashed, mark=none,domain=-1.1:-1,samples=100,yticklabels=\empty] {0.1*cos(deg(x*3.141593))+ 0.1*cos(deg(2*x*3.141593)) + 0.1*sin(deg(5*x*3.141593))+0.5};
    
            \path[name path=axis1] (axis cs:-1,0) -- (axis cs:-0.125,0);
    
    			\addplot[mark='',only marks] coordinates{(0.-1,-0.01)}node[anchor=north]{\scriptsize{$p_{\zeta-2}$}};
    			\addplot[mark='',only marks] coordinates{(0.-0.125,-0.01)}node[anchor=north]{\scriptsize{$p_{\zeta-1}$}};				
                \addplot [fill=blue, fill opacity=0.05] fill between[of=f and axis1, soft clip={domain=-1:-0.125},];
                \addplot[mark='',only marks] coordinates{(-0.5,0.01)}node[anchor=south]{\scriptsize{$M_{\zeta-1,i}(t)$}};
        		\draw[dotted] (axis cs:-1,1.25) -- (axis cs:-1,-1.25);
    			\draw[dotted] (axis cs:-0.125,1.25) -- (axis cs:-0.125,-1.25);
    
            \path[name path=axis2] (axis cs:-0.125,0) -- (axis cs:0.5,0);
    
    			\addplot[mark='',only marks] coordinates{(0.5,-0.01)}node[anchor=north]{\scriptsize{$p_\zeta$}};
    			\addplot [fill=red, fill opacity=0.05] fill between[of=f and axis2, soft clip={domain=-0.125:0.5},];
                \addplot[mark='',only marks] coordinates{(0.2,0.01)}node[anchor=south]{\scriptsize{$M_{\zeta, i}(t)$}};
    			\draw[dotted] (axis cs:-0.125,1.25) -- (axis cs:-0.125,-1.25);
    			\draw[dotted] (axis cs:0.5,1.25) -- (axis cs:0.5,-1.25);
    
            \path[name path=axis3] (axis cs:0.5,0) -- (axis cs:1,0);
    
    			\addplot[mark='',only marks] coordinates{(1,-0.01)}node[anchor=north]{\scriptsize{$p_{\zeta+1}$}};
    			\addplot [fill=black, fill opacity=0.05] fill between[of=f and axis3, soft clip={domain=0.5:1},];
                \addplot[mark='',only marks] coordinates{(0.75,0.01)}node[anchor=south]{\scriptsize{$M_{\zeta +1, i}(t)$}};
    			\draw[dotted] (axis cs:0.5,1.25) -- (axis cs:0.5,-1.25);
    			\draw[dotted] (axis cs:1,1.25) -- (axis cs:1,-1.25);

		    \end{axis}
        \end{tikzpicture}
    \end{center}
    \caption{Schematic representation of influence masses for a given density function, $f_i(w,t)$, at some time $t$.} \label{fig:inflmass}
    \end{figure}

    We now discuss the nature of our leader species, in terms of political parties. Let there be $\mc{P}$ parties, $\Pi_1, ..., \Pi_\mc{P}$, in our political system, and $a=p_0 < p_1<...<p_{\mc{P}-1}< p_{\mc{P}} = b$ be a partition of $\I=[a,b]$. Then we distribute the leader species population in the intervals $\I_\zeta = [p_{\zeta-1}, p_\zeta]$, which we call ``influence intervals". We run an ``election'' at time t, on the microscopic level, by assuming that all individuals vote for the party who controls the influence interval they are contained within, and on the mesoscopic and macroscopic levels, by computing ``influence masses",
    $$M_{\zeta, i}(t) = \int_{\I_\zeta} f_\zeta(w,t) \ \od w$$
    for every $\zeta = 1,...,\mc{P}$, a graphical representation of which is presented in Figure \ref{fig:inflmass}.

    The voter intention polling data we use is organised into tables by characteristics, with party on the vertical axis and the groups (e.g. leave and remain for 2016 Brexit Vote, 18-24 year-old etc. for age) on the horizontal axis. A table for an example characteristic $k$ is displayed in Table \ref{tab:PFunction:examplechar} and the actual table for the 2016 Brexit vote is displayed in Table \ref{tab:PFunction:brexitchar}.

    \begin{table}[h] 
    \centering
    \caption{Example of how the characteristic data is presented.}
    \renewcommand{\arraystretch}{1.2}
        \begin{tabular}{c|cccc}
            Characteristic $k$ & Group 1 & Group 2 & $\cdots$ & Group $m$\\
            \hline
            $\Pi_1$ & $x_{1,1,k}$ & $x_{1,2,k}$ & $\cdots$ & $x_{1,m,k}$ \\
            $\Pi_2$  & $x_{2,1,k}$ & $x_{2,2,k}$ & $\cdots$ & $x_{2,m,k}$ \\
            $\vdots$ &  $\vdots$ & $\vdots$  & $\ddots$ & $\vdots$  \\
            $\Pi_\mc{P}$  & $x_{\mc{P},1,k}$ & $x_{\mc{P},2,k}$ & $\cdots$ & $x_{\mc{P},m,k}$ 
        \end{tabular}
        \label{tab:PFunction:examplechar}
    \end{table}

    \begin{table}[h]
    \centering
    \caption{The voter intention data for 2016 Brexit vote from \cite{bib:YouGov:voterintention}.}
    \renewcommand{\arraystretch}{1.2}
        \begin{tabular}{r|cc}
            Brexit Vote & Leave & Remain \\
            \hline
            Labour & 0.1313 & 0.4081 \\ 
            Green & 0.0202 & 0.0612 \\
            LibDems & 0.0404 & 0.2551 \\
            Other & 0.0303 & 0.0306 \\
            Conservative &0.6969 & 0.1938 \\
            Brexit & 0.0606 & 0 
        \end{tabular}
        \label{tab:PFunction:brexitchar}
    \end{table}

    For each binary interaction \eqref{int:LL}-\eqref{int:cc} we can find the influence intervals $\I_\zeta$ to which $w,v$ belong to, assume that $w\in \I_\mu$ and $v\in \I_\nu$, for some $\mu, \nu = 1,...,\mc{P}$. We look up the relevant values in the tables to get the values of $x_{\mu,i,k}$ and $x_{\nu,j,k}$. We want our characteristic comparison function $P_{char}$ to represent the idea of net movement towards more popular opinions, popular in the sense of voter intentions from the pre-election polls. That is individuals with popular opinions move slower towards the unpopular opinions than the individuals with unpopular opinions move towards the more popular opinion. To this end, we let our characteristic comparison function depend on $x_{\mu,i,k} - x_{\nu,j,k}$, which gives a large, positive value when the party, in whose influence the individual resides, is more popular (i.e. $x_{\mu,i,k}$ dominates) and a large negative value when the party, in whose influse the other individual resides, is more popular (i.e. $x_{\nu,j,k}$ dominates). When both parties are equally popular, this value will be zero and therefore will not have a bias towards one party or the other. We are therefore using $x_{\mu,i,k} - x_{\nu,j,k}$ since it is representative of the popularity of the individual's opinion.
    
    The value of $x_{\mu,i,k} - x_{\nu,j,k}$ is in $[-2,2]$, but as previously stated our $P$-function's value must be in $[0,1]$, hence we project the value of $x_{\mu,i,k} - x_{\nu,j,k}$ into $[0,1]$. To achieve this we use the linear function, $$H(x) = \frac{1}{4}(2+x).$$ Since we are using $x_{\mu,i,k} - x_{\nu,j,k}$ as a measure of the popularity of the individual's opinion $w$, we want $P_{char}$ to be decreasing with respect to its argument $x_{\mu,i,k} - x_{\nu,j,k}$, to model the decrease in movement from, and the increase in movement to, popular opinions. To this end we use a second function, $$G(x) = \bigg(1-x^a\bigg)^b,$$ for some $a,b>0$. Note that $G$ has the properties: (a) $G(0) = 1$ and $G(1)=0$; and (b) this function is decreasing in $[0,1]$, for any value of $a,b>0$. These properties, in combination, lead to the value of $G$ being close to zero when $x_{\mu,i,k}$ is dominant (or the individual holds a more popular opinion), and represents a small movement towards the other more unpopular opinion and vice versa. Finally, we compose these functions and thus we use the function, $$F(x)= G(H(x))= \bigg(1-\left(\frac{1}{4}(2+x)\right)^a\bigg)^b,$$ with $a,b > 0$. Thus our $P_{char} $ function is, $$P_{char}(w,v;c_{i,k}, c_{j,k}) = F(x_{\mu,i,k} - x_{\nu,j,k}),$$ when viewed on the microscopic level. In general, we can write this function as the sum of characteristic functions which sort $v,w$ into their respective influence intervals as follows;
    \begin{equation}
        P_{char}(w,v;c_{i,k}, c_{j,k}) = \sum^\mc{P}_{\mu, \nu = 1} \mathbbm{1}_{w \in [p_{\mu-1}, p_\mu]} \mathbbm{1}_{v \in [p_{\nu-1}, p_\nu]}  F(x_{\mu,i,k}-x_{\nu,j,k}) \label{eq:ourP}
    \end{equation}
    where $\mathbbm{1}_{x\in A}$ denotes the characteristic function for the set $A$.  

\section{Numerical experiments} \label{sec:Numerical}

    
    We separate the analysis of the methods into two types, the Monte Carlo simulation of the Boltzmann-type system \eqref{eq:Boltz}--\eqref{eq:BoltzL}, and the Finite Element Method for the Fokker-Plank system \eqref{eq:Fokker}--\eqref{eq:FokkerL}. 
    
    Throughout this section, we have chosen $\eta_i$ all independent and identically distributed, in such a way that they can only attain two values, $\pm 0.01$, with equal probability, to ensure that the interactions remain in $\I$. Unless otherwise stated, we use the values of $a=2$ and $b=8$ in the $P_{char}$ function in Section~\ref{sec:Deriv}.\ref{sec:Deriv:PFunction} since these have offered the best results, $\alpha_{i} = \beta_{i} = 1$ for all $i = 1, 2, ..., N, L$. We construct our initial densities from available polling data and from population values for the specific region we wish to run simulations in. We do this by first finding the influence mass of the Party $\Pi_\zeta$ for a given species $i$, multiplying the relevant columns of our data tables, which, for a species whose members have characteristic $k$ and are in Group $\xi$, is the $\xi$-th column from Table \ref{tab:PFunction:examplechar} and the relevant column from every characteristics table, element-wise. We will call the elements of the resulting column $X_{\zeta,i}$. Recall that by definition 
    $$M_i = \sum_{\zeta}M_{\zeta,i},$$
    where we recall that $M_{\zeta,i}$ are the influence masses from Section~\ref{sec:Deriv}.\ref{sec:Deriv:PFunction}. Given some distribution of mass between the follower species $M_i$, we can compute the initial influence masses for the parties in  
    $$M_{\zeta, i} = \frac{v_\zeta X_{\zeta,i}}{\sum_{\zeta} v_\zeta X_{\zeta, i}}  M_i,$$
    with $v_\zeta$ being the percentage of votes allocated to $\Pi_\zeta$ in some test election, either through the recorded results of some previous election or some pre-election polling data. We define the initial densities of the species as a weighted sum of characteristic functions in the following way
    $$f_i(w,0) = \sum_\zeta M_{\zeta,i} \mathbbm{1}_{w\in \I_\zeta},$$
    for all $\zeta = 1,2,...,N$.
    
    The initial density of the leader species is chosen as a weighted sum of delta distributions,
    \begin{equation} \label{eq:initiallead}
        f_L(w,0) = \sum_{k=1}^6 p_k\delta(w-L_k),
    \end{equation}
    with $p_k$ representing the percentage of the vote that each party attained in this, or another, test election, and $L_k$ representing the party locations. In all the simulations presented here, we set the mass of the leader species to have 5\% the total mass of the follower species, in a similar way to \textit{D\"uring et al} \cite{bib:During:strongleaders}. For purposes of the discretisation in the Fokker-Plank FEM simulation, we approximated the delta functions in $f_L(w,0)$ with the scaled mollifier functions
    \begin{equation*}
        \phi(x) = \begin{dcases}
        \exp\left(-\frac{({x}/{\varepsilon})^2}{1-({x}/{\varepsilon})^2}\right), & \text{for } x \in [-1,1], \\
        0, & \text{otherwise},
                   \end{dcases}
    \end{equation*}
    with scale factor $\varepsilon \ll 1$.


    \subsection{Monte-Carlo simulations of the Boltzmann-type system} \label{sec:Numerical:MCS}

    
    We perform a series of kinetic Monte Carlo simulations for the Boltzmann-type system \eqref{eq:Boltz}--\eqref{eq:BoltzL}. In these simulations, called Direct Simulation Monte Carlo (DSMC) or Bird's scheme, we choose individuals from the population non-exclusively, and perform the interaction rules \eqref{int:cc}, \eqref{int:cL} and \eqref{int:LL}, depending on the species those individuals fell into. We run all simulations with $M=10500$ total individuals, with 500 individuals in the Leader species and the remaining $10000$ individuals distributed into follower species. We distribute these individuals using data from previous elections and population data.
    
    Each time step in our simulation corresponds to $M$ interactions and we run the simulation for 1000 time steps averaging over the last 50 time steps. We average the results of eight realisations to give a good approximation of the long-term density. We discretise the spatial domain $\I = [-1,1]$ in 50 sub-intervals.

    
    \subsection{Finite Element discretisation of the Fokker-Planck system} \label{sec:Numerical:FPS}

    Throughout this section, we use $(f,g)$ to indicate the $L^2$ inner product of the function $f$ and $g$, that is, $(f,g)=\int_\I f(w)\cdot g(w) \ \od w.$

    We discretise our time axis with $A=300$ nodal points, and space axis with $B=100$ nodal points. We write the nodal values of our functions $f_i(w_k,t^n) = f_{i,k}^n$ and the vector of every spatial nodal value, at time $t^n>0$, as $\mb{f}_i^n$. We then use a semi-implicit Euler method to discretise our coupled system of non-linear, non-local PDEs in time and a finite element method (FEM) for discretisation in space, with quadratic, Lagrangian basis functions $b_k \in C_c(\I;\real)$ for $k = 1,2,...,B $, which leads to the following linear system:
    \begin{equation} \label{FPM:discrete}
            \sum_{i=1}^N \bigg((\mb{M} + \mb{U}_{i} - \mb{C}^n_{Li}) \mb{f}^{n+1}_i + \sum_{j=1}^N (\mb{V}_{ij} - \mb{T}_{ij}^n) \mb{f}_j^{n+1}\bigg) = \sum_{i=1}^N \mb{M} \mb{f}^n_i,
    \end{equation}
     where $\mb{M} = (b_k,b_l)_{(k,l)}$ is the mass matrix, $\mb{C}^n_{Li}$ and $\mb{T}_{ij}^n$ are the discretised transport matrices,
given by
    $$\mb{C}^n_{Li} = 
    \Delta t  \left(\left[\frac{1}{2\tau_{iL}}K_{L,iL}^n + \sum_{j=1}^N \frac{1}{2\tau_{ij} }K_{j,ij}^n\right]b_k, b_l'\right)_{(k,l)},\qquad 
    \mb{T}_{ij}^n = \Delta t \left(\frac{\alpha_{ij}}{2\tau_{ij}} K_{i,ij}^n b_k, b_l'\right)_{(k,l)},$$

    \noindent with $K_{L,iL}^n(w)$ and $K_{m,ij}^n(w)$ as the midpoint rule approximation of the convolution operators, $\mc{K}_{m,ij} (w, t^n)$ evaluated at the previous time step, in the transport terms of the Fokker-Planck equations and $\mb{U}_i$ and $\mb{V}_{ij}$ are the discretised Laplacian term matrices,
    $$\mb{U}_i = \Delta t \left(\left[\frac{\lambda_i M_L}{4\tau_{iL}} + \sum_{j=1}^N \frac{\lambda_i M_j}{4\tau_{ij}}\right]D^2 b_k',b_l'\right)_{(k,l)},\qquad
    \mb{V}_{ij} = \Delta t \left(\frac{\lambda_i M_j \beta^2_{ij}}{4\tau_{ij}\alpha_{ij}}D^2 b_k', b_l'\right)_{(k,l)}.$$

    
    \subsection{Application to the 2019 General Election in the UK} \label{sec:Numerical:DataUsage}


    We now consider an application of the above model on the United Kingdom General Election in 2019. The electoral areas of the United Kingdom are various local areas called constituencies, each of which elect a single member of parliament (MP) to the House of Commons by a first-past-the-post system, that is the candidate with the most votes in the constituency is declared the winner and becomes the MP. The political party with the largest number of seats in the House of Commons is, by convention, called the winning party and the Leader of the winning party is usually chosen to be the Prime Minister. We want to run simulations of this model on an individual constituency level, however the freely available voter intention data from YouGov we use \cite{bib:YouGov:voterintention} is presented on a national level for  England, Scotland and Wales. We therefore approximate local data using the methods presented in this section. We choose to run simulations in various regions, 1) the Office of National Statistics' (ONS) regions of England, which are: the East of England, East Midlands, London, North East, North West, South East, South West, West Midlands, and Yorkshire and the Humber; and 2) the county of Nottinghamshire containing the constituencies Ashfield, Bassetlaw, Broxtowe, Gedling, Mansfield, Newark, Nottingham East, North and South, Rushcliffe, and Sherwood. We use the ONS' regions since they allow for results representative of the overall result in these large areas and the county of Nottinghamshire because Ashfield, Bassetlaw and Mansfield are part of the so-called ``Red Wall", a group of constituencies in the North and Midlands of the UK who traditionally elected Labour MPs, yet elected many Conservative MPs in the 2019 General Election, in what was considered to be a decisive factor in the overall outcome of the election \cite{bib:Guardian, bib:BBC}. We use the 2011 Census of the UK \cite{bib:ONS:2011census} to get the total population and the demographic population for the above areas. 
    
    We have made the assumption that supporters of the UK independence party (UKIP) from 2015 and 2017 General Elections \cite{bib:HOC:2015GE, bib:HOC:2017GE} are aligned closer to the political views of the Brexit Party in the 2019 General Election. This is because the manifesto of the Brexit Party is very similar to that of UKIP in those years. In the voter intention data from YouGov, the parties of Plaid Cymru and the Scottish National Party (SNP) are included but since they do not run a candidate in any constituency in the East of England, nor in Nottinghamshire, we have omitted them from the simulations.
    
    For these simulations, we focus on two characteristics, age, split into 18--24 year-old, 25--49 year-old, 50--64 year-old and 65+ year-old, and self-reported 2016 Brexit Vote, split into Leave and Remain. Thus our eight follower species will be 18-24 year-olds, who voted leave, 25-49 year-olds who voted leave etc. We again reiterate that since we are modelling this process over a number of years, there will be some demographic shift as individuals at the upper end of an age range will shift to a different species, however it may still be reasonable not to take this into account here, if these individuals do not exhibit a large change in political view over this period. If required, demographic changes (migration to other species, birth/death) could be incorporated by adding gain and loss terms similarly as in \cite{bib:During:strongleaders}.
    
    We choose our party locations, that is the values of $L_k$ in \eqref{eq:initiallead}, to be: Labour at $-0.7$, Green at $-0.4$, Liberal Democrats (LibDems) at $-0.1$, Other at $0.1$, Conservative at $0.4$, and the Brexit Party at $0.7$. This leads us to an intuitive way to choose the influence intervals, $\I_\zeta$, by choosing the midpoint between the party locations as the partitions, as well as $-1$ and $1$ for the most extreme opinions: Labour has influence in the region $[-1, -0.55)$, Green in $[-0.55, -0.25)$, LibDems in $[-0.25,0)$, Other in $[0, 0.25)$, Conservatives in $[0.25, 0.55)$, and the Brexit Party in $[0.55,1]$. In the FEM of the Fokker-Planck system, we will use the scale factors for our mollifier functions as $\varepsilon = 0.08$, so that the peaks that are associated with the parties remains entirely within that parties influence interval.

    \subsubsection{Case study: East of England} \label{sec:Data:EOECaseStudy}
    
    In this subsection, we look specifically at the East of England region. However, every step we use in this section, can be applied similarly to all areas described above. 

    To process the data, we take the results for the 2017 General Election \cite{bib:HOC:2017GE} and compute the total vote tallies for the East of England. This gives us the popular vote percentages in the region, the result of which is displayed in Table \ref{tab:PVP}. 
    
    \begin{table}[h]
    \centering
    \caption{Popular vote percentage for the East of England for the 2017 General Election. }
    \renewcommand{\arraystretch}{1.2}
        \begin{tabular}{r|c}
             & proportion of vote \\
            \hline
            Labour & 0.3271 \\ 
            Green & 0.0189 \\
            LibDems & 0.0788 \\
            Other & 0.0034 \\
            Conservative & 0.5464 \\
            UKIP & 0.0251
        \end{tabular}
        \label{tab:PVP}
    \end{table}
    
\begin{figure}
	    \begin{subfigure}[b]{0.5\linewidth}
		    \includegraphics[scale=0.4]{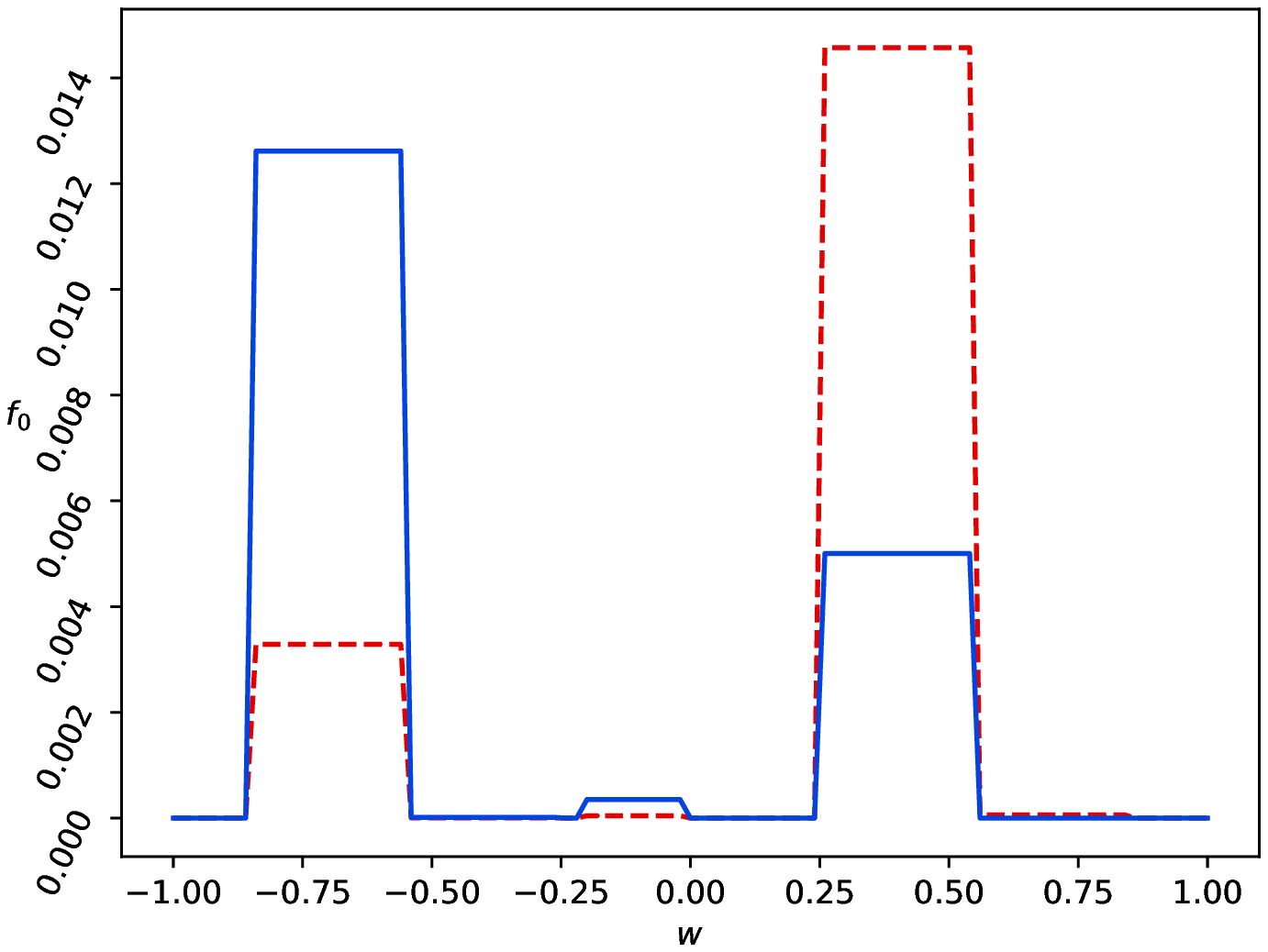}
		    \caption{18--24 year-old}
	    \end{subfigure}
		\hspace{0.04cm}
	    \begin{subfigure}[b]{0.5\linewidth}
		    \includegraphics[scale=0.4]{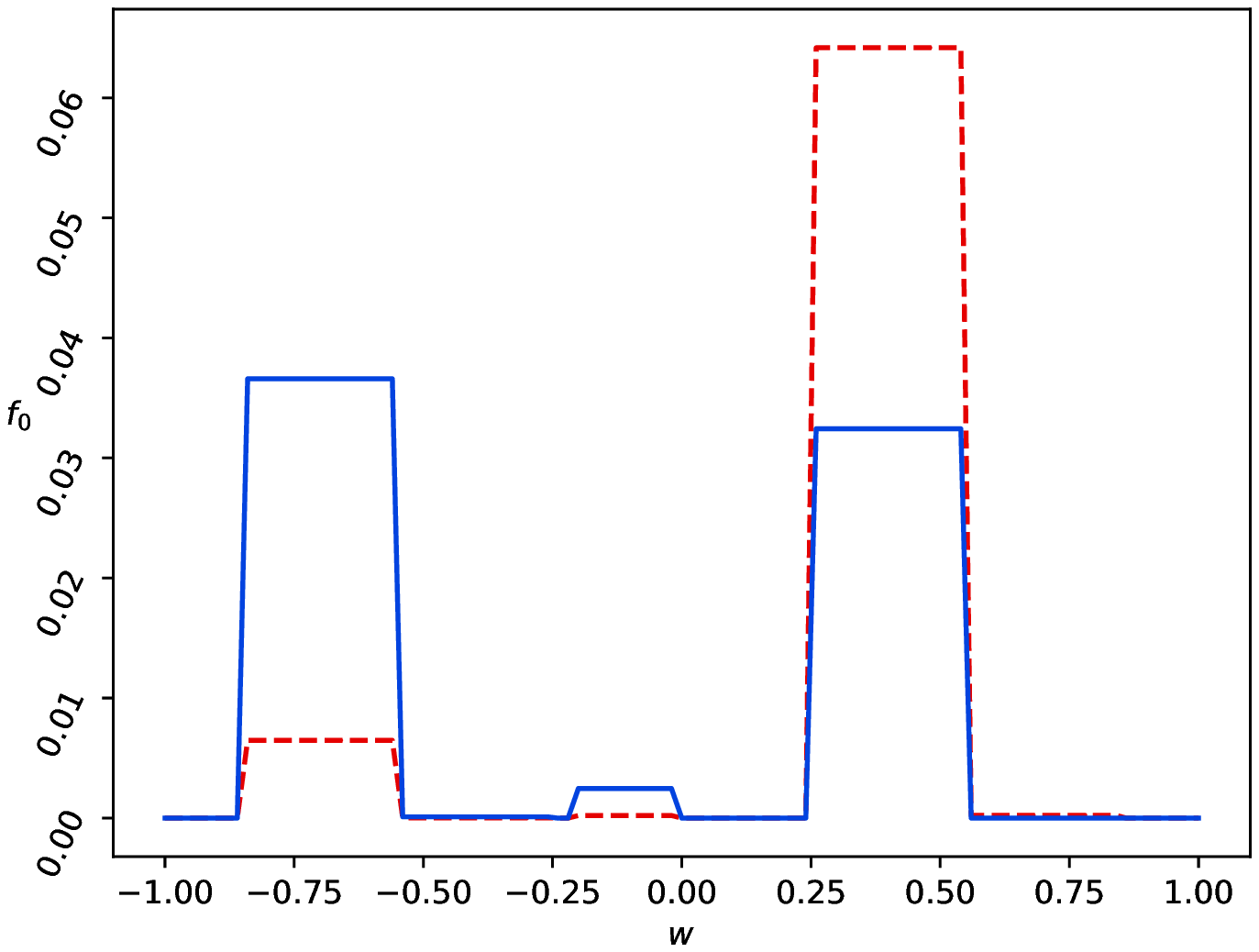}
		    \caption{25--49 year-old}
	    \end{subfigure} 
    
    	\begin{subfigure}[b]{0.5\linewidth}
	    	\includegraphics[scale=0.4]{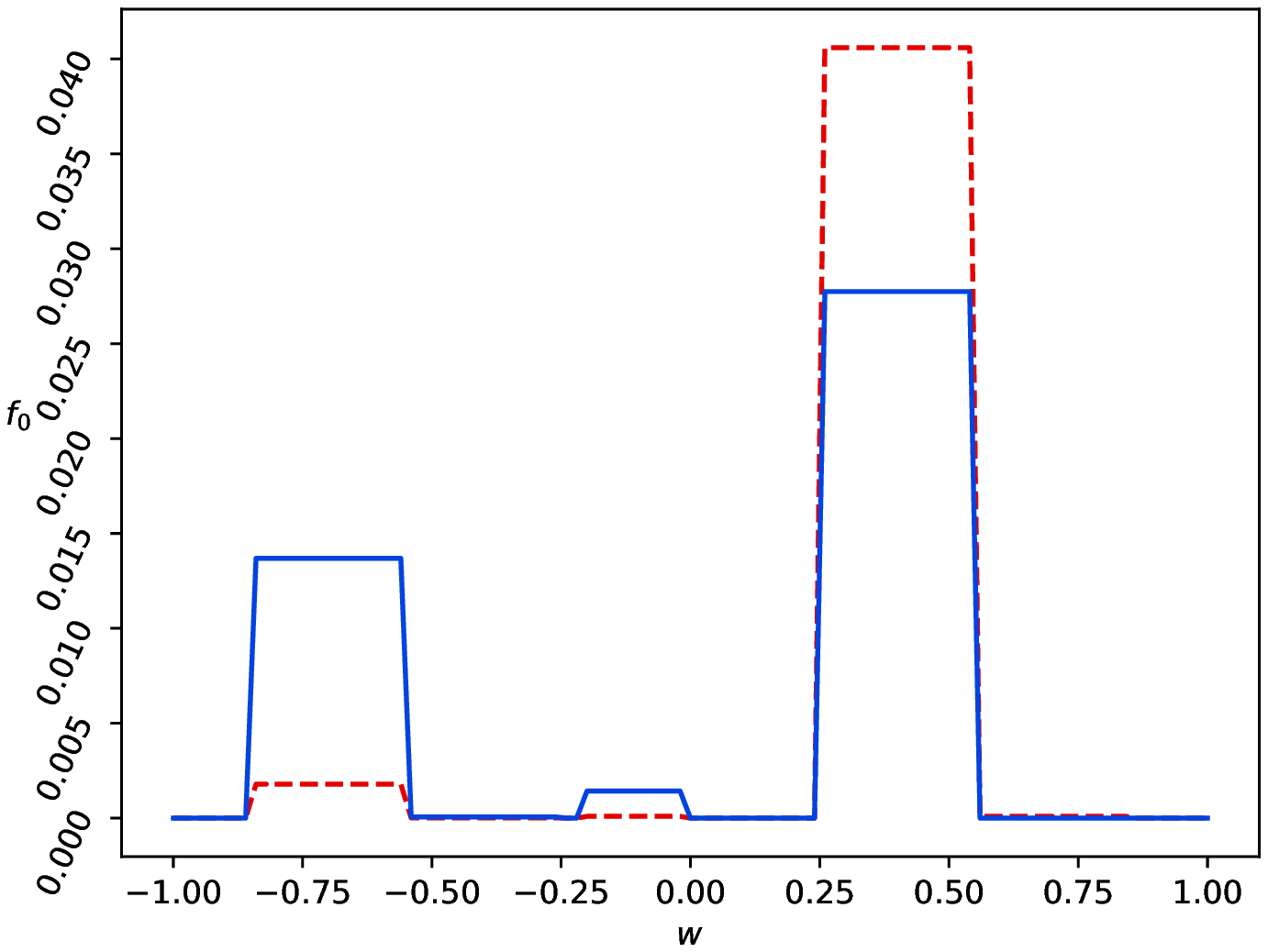}
		    \caption{50--64 year-old}
	    \end{subfigure} 
	    \hspace{0.04cm}
	    \begin{subfigure}[b]{0.5\linewidth}
		    \includegraphics[scale=0.4]{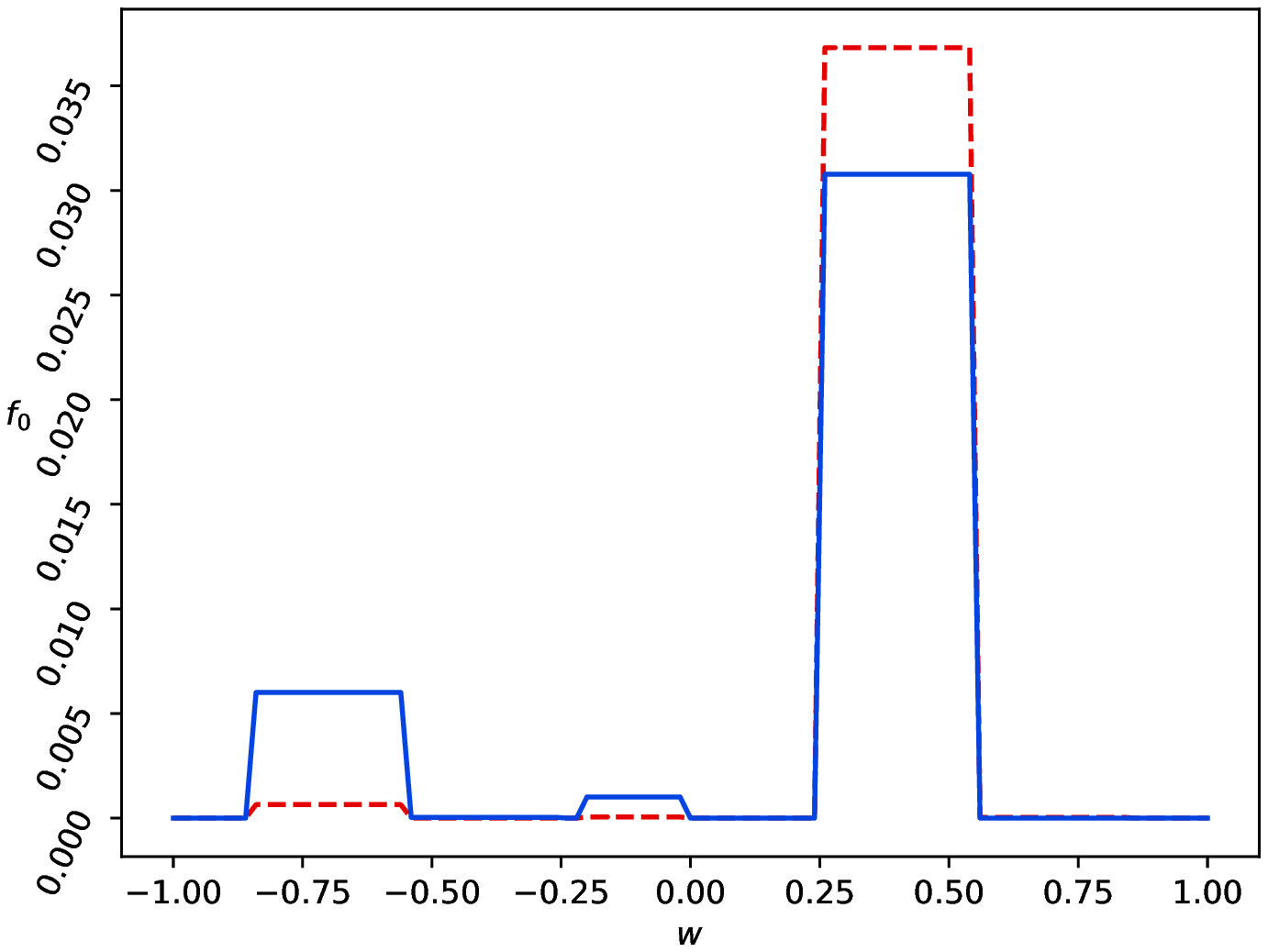}
            \caption{65+ year-old}
	    \end{subfigure}
	    \caption{Initial follower densities for the eight follower species, from 2015 General Election Results. Each graph shows the initial density for the leave (solid blue) and the remain (dashed red) characteristic of an age group.}	
	    \label{fig:4.0intro:ICs}
    \end{figure}

    We then use the demographic voter intention data from YouGov \cite{bib:YouGov:voterintention}, and multiply this by the previous election result of the region and normalise it as seen in Table \ref{tab:RegDat}.
    \begin{table}[h]
    \centering
    \caption{YouGov data (left) and regionalised data (right) for the Brexit vote for the East of England.}
    \renewcommand{\arraystretch}{1.2}
        \begin{tabular}{r|cc|cc}
             & \multicolumn{2}{c|}{YouGov national} & \multicolumn{2}{c}{East regionalised} \\
             & Leave & Remain & Leave & Remain\\
            \hline
            Labour & 0.1313 & 0.4081 & 0.1001 & 0.5119 \\ 
            Green & 0.0202 & 0.0612 & 0.0009 & 0.0044 \\
            LibDems & 0.0404 & 0.2551 & 0.0074 & 0.0771 \\
            Other & 0.0303 & 0.0306 & 0.0002 & 0.0004 \\
            Conservative &0.6969 & 0.1938 & 0.8877 & 0.4061 \\
            Brexit & 0.0606 & 0 & 0.0035 & 0 
        \end{tabular}
        \label{tab:RegDat}
    \end{table}
    We compute the masses of the species densities $M_i$ to reflect the population size in the 2011 Census which is already split by region and age, then divide them based on the result of 2016 European Union Referendum in the East of England region, the data for which is freely available from the House of Commons Library \cite{bib:ElectoralComission:2016Brexit}. For example, to compute the approximate number of 18-24 year-olds who voted to leave, we find the percentage of the electorate of the East of England that is 18-24 years-old, $$\frac{\text{\# of 18-24 year-olds}}{\text{\# of eligible voters in EOE}} = \frac{490,197}{4,617,534} = 0.1061,$$
    then we compute the percentage of the electorate that voted to leave in the 2016 Brexit vote in the East of England, $$\frac{\text{\# of voters that voted leave}}{\text{\# of valid votes cast}} = \frac{1880367}{3328983} = 0.5648.$$ We approximate the mass of the population that is 18-24 years-old and  voted leave by taking the product of the above values, that is $0.1061 \times 0.5648 = 0.0599$. We repeat this for all species, and we obtain the values in Table \ref{tab:masses}. We use the method from Section \ref{sec:Numerical}, using the voter intention data from YouGov and the 2015 General Election Results to generate our initial conditions for the follower species, as described in Section~\ref{sec:Numerical}. A graphical representation of these initial functions $f_i(w,0)$ is given in Figure \ref{fig:4.0intro:ICs} where we have displayed the eight initial follower densities by age in subgraphs and by self-reported Brexit vote by colour. 
    
\begin{figure} 	
        \centering
	    \includegraphics[scale=0.5]{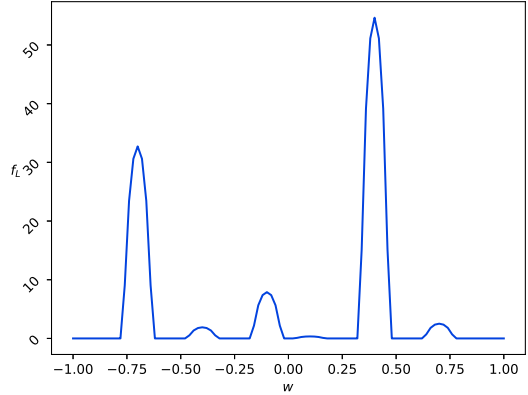}
	    \caption{The mollified, initial Leader species density for the East of England Region.}
	    \label{fig:4.0intro:Leaders}
    \end{figure}

    For the leader species initial condition, we use directly the results from the 2015 General Election and the graph of the resultant initial function, $f_i(w,0)$ in the East of England Region used in the Fokker-Plank simulation is presented in Figure \ref{fig:4.0intro:Leaders}.
    
    \begin{table}[h]
    \centering
    \caption{The masses $M_i$ of the follower species $i$ in the East of England.}
    \renewcommand{\arraystretch}{1.2}
        \begin{tabular}{r|cc}
             & Leave & Remain \\
            \hline
            18-24 & 0.0599 & 0.0461 \\ 
            25-49 & 0.2374 & 0.1829 \\
            50-64 & 0.1421 & 0.1094 \\
            65+ & 0.1252 & 0.0965
        \end{tabular}
        
        \label{tab:masses}
    \end{table}
    
    
    \subsection{Discussion of results}


    We present here the results from our numerical experiments of the model presented thus far. First we look at the how the DSMC and FEM methods compare when applied to the East of England, as described in Section~\ref{sec:Numerical}.\ref{sec:Numerical:DataUsage}, and lastly in this section, we investigate the effect of the $P_{char}$ function on the model by comparing the choice from Section~\ref{sec:Deriv}.\ref{sec:Deriv:PFunction} to a model analogous with that of \textit{D\"uring et al.}. We use the FEniCS Project 
    \cite{LoggMardalEtAl2012a,AlnaesBlechta2015a}, to simulate the Fokker-Plank system, with graphs drawn using MatPlotLib \cite{matplotlib} and the DSMC experiments are run using Python 3. The parameters used for both models are similar to those used in \cite{bib:During:strongleaders, bib:During:elo}. In the Fokker-Planck system larger (smaller) values of $\lambda_i$ do smooth out profiles more (less),  but do not affect the positions of local maxima. The choices of $\gamma_i$ and $\eta$ affect the results of the DSMC simulation in a similar way (since $\lambda_i=\sigma_i^2/\gamma_i$ in the limit process in Section~\ref{sec:Deriv}.\ref{sec:Deriv:BoltFokker}. The choice of $r$ determines the range of interactions, and hence smaller values of $r$ lead to a larger number of local maxima, compare the results in \cite{zbMATH07146312}.
    
    \subsubsection{Numerical Results: comparing Boltzmann-type to Fokker-Planck}
    
    First, we run the DSMC experiments for the Boltzmann-type system and for the Fokker-Planck system with the same -- or equivalent -- parameters. We do this to illustrate that the numerical results from the Boltzmann-type system and the Fokker-Planck system are good approximations of each other. We choose our parameters in the simulations as: $r=0.5$, $\gamma_i = \gamma = 0.3$, $\eta = \pm 0.01$ each with probability ${1}/{2}$, $\gamma_L= 0$ and $D_L(w) \equiv 0$ in the DSMC for the Boltzmann-type system; and $r = 0.5$, $\tau_{iL} = 0.05$, $\tau_{i,j} = 2.5$, $\lambda_i = 0.033$, $\Delta s = 0.1$ in the FEM for the Fokker-Planck system.
    
    \begin{figure}
	    \begin{subfigure}[b]{0.5\linewidth}
		    \includegraphics[scale=0.4]{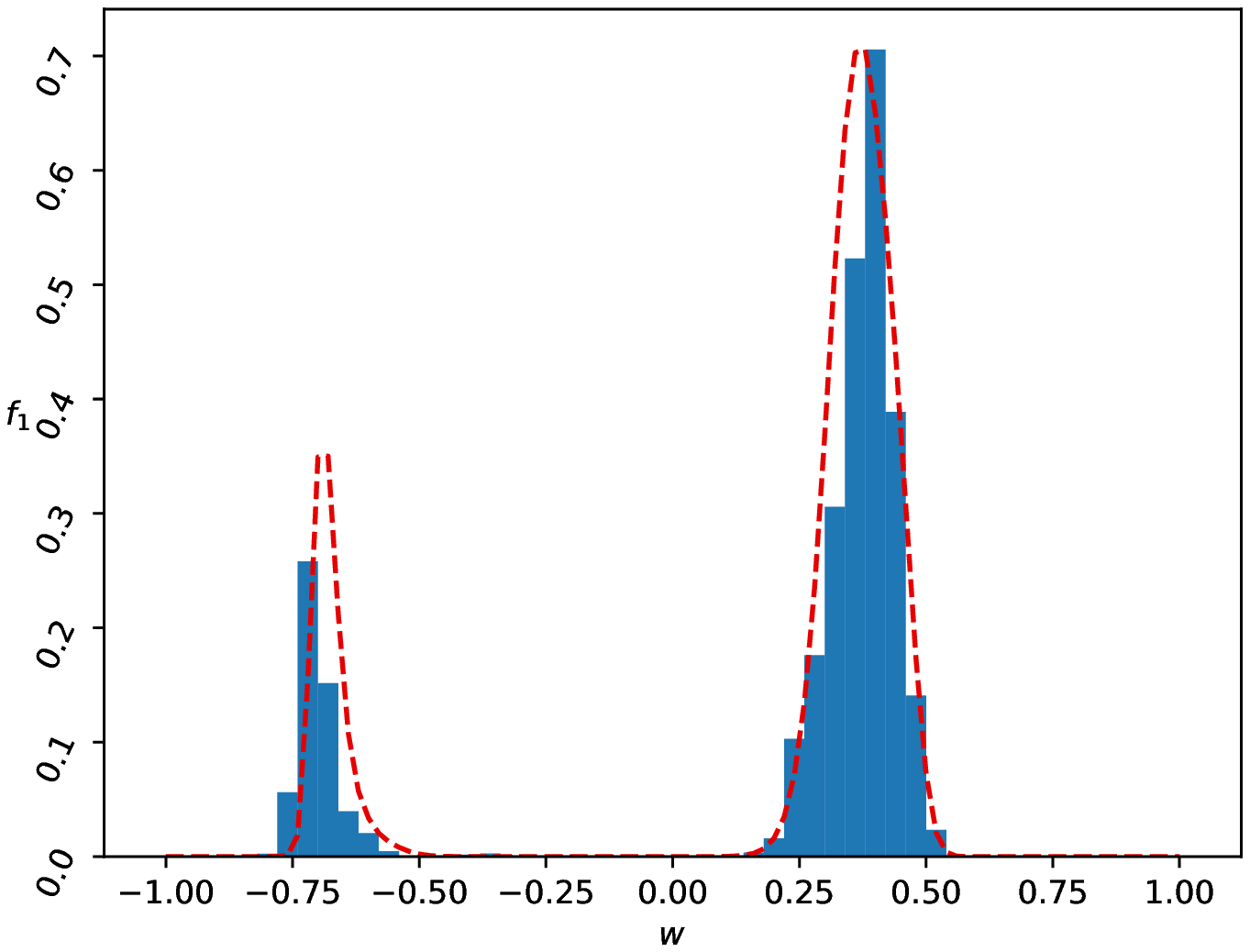}
		    \caption{18--24 year-old, leave}
	    \end{subfigure}
		\hspace{0.04cm}
	    \begin{subfigure}[b]{0.5\linewidth}
		    \includegraphics[scale=0.4]{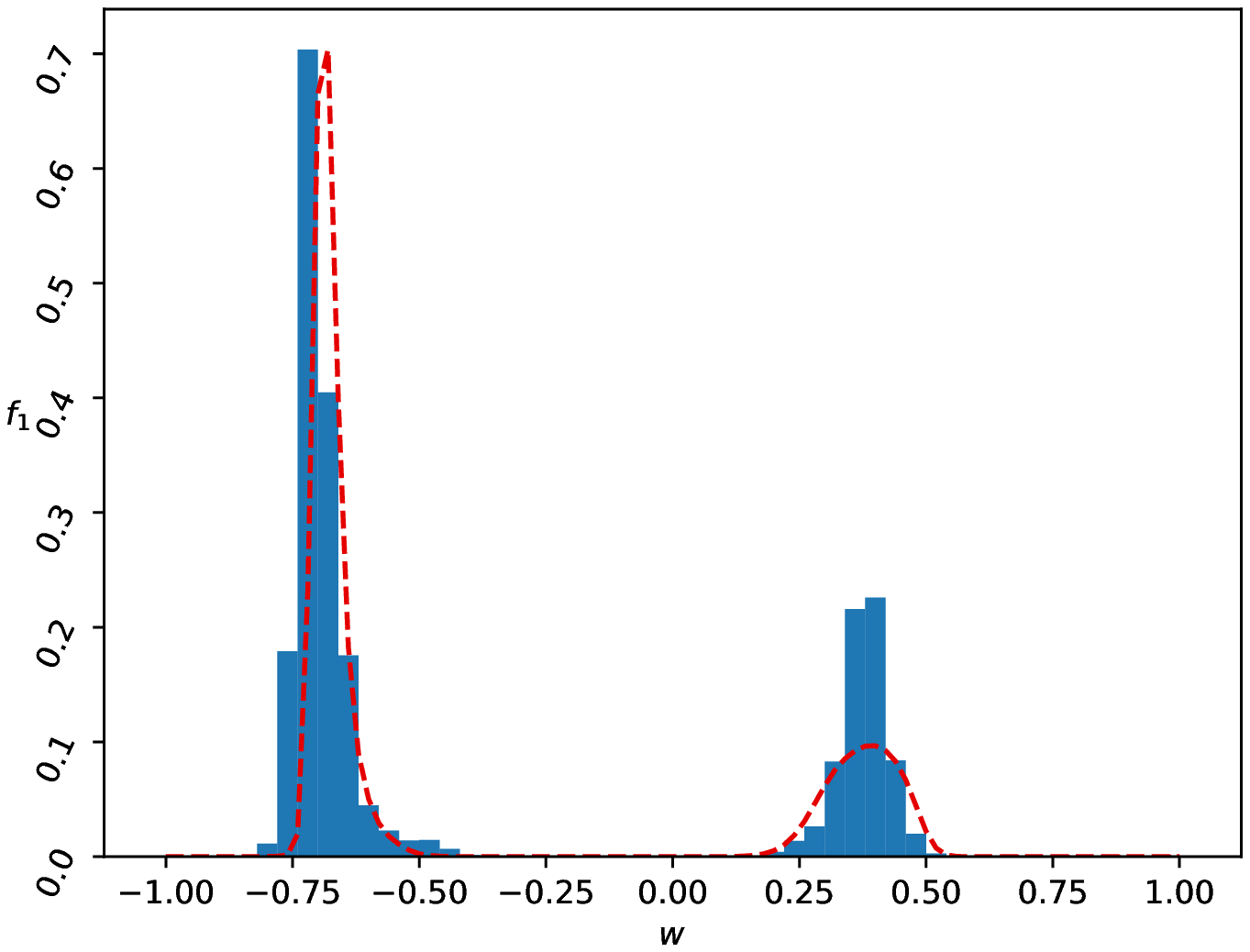}
		    \caption{18--24 year-old, remain}
	    \end{subfigure} 
    
    	\begin{subfigure}[b]{0.5\linewidth}
	    	\includegraphics[scale=0.4]{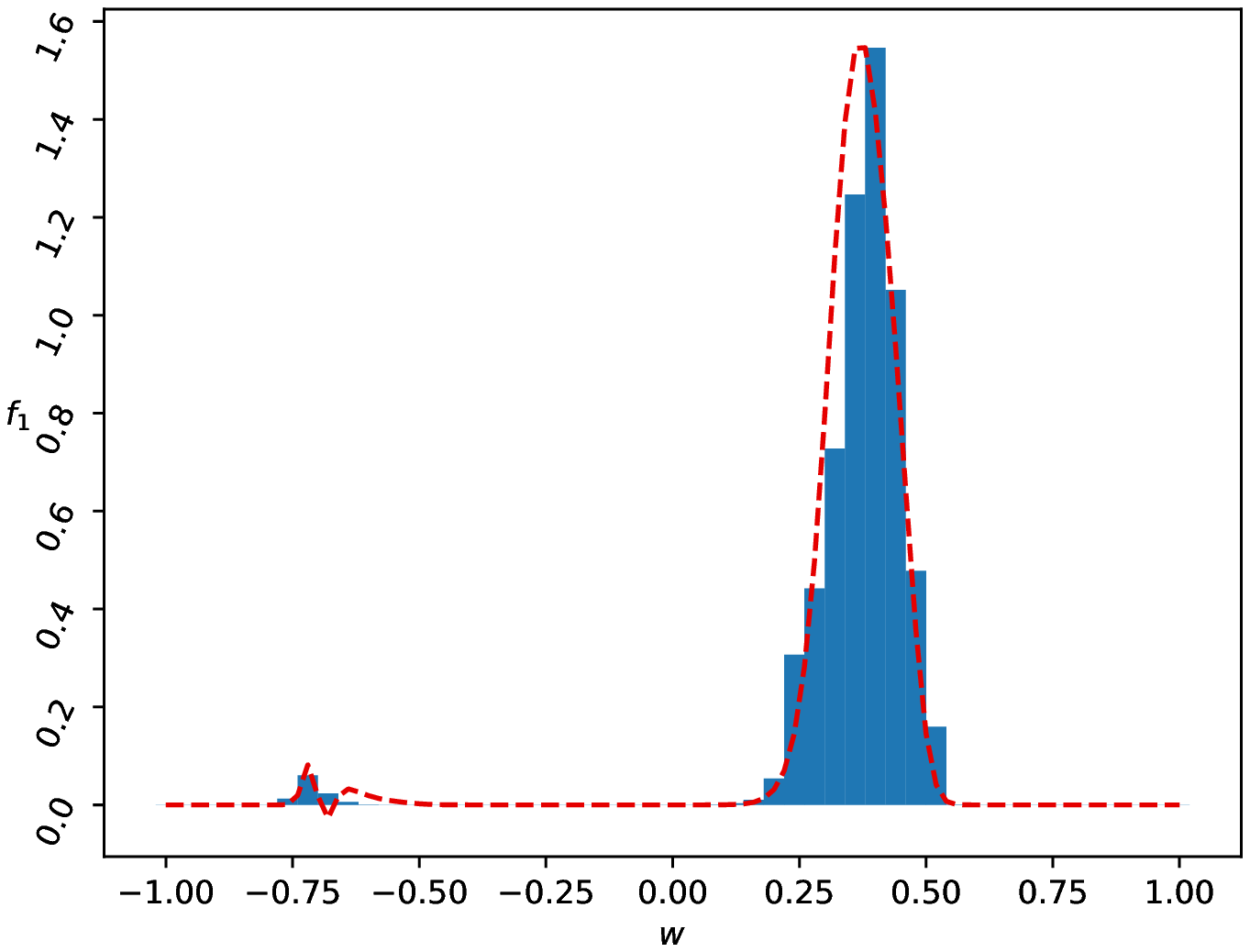}
		    \caption{65+ year-old, leave}
	    \end{subfigure} 
	    \hspace{0.04cm}
	    \begin{subfigure}[b]{0.5\linewidth}
		    \includegraphics[scale=0.4]{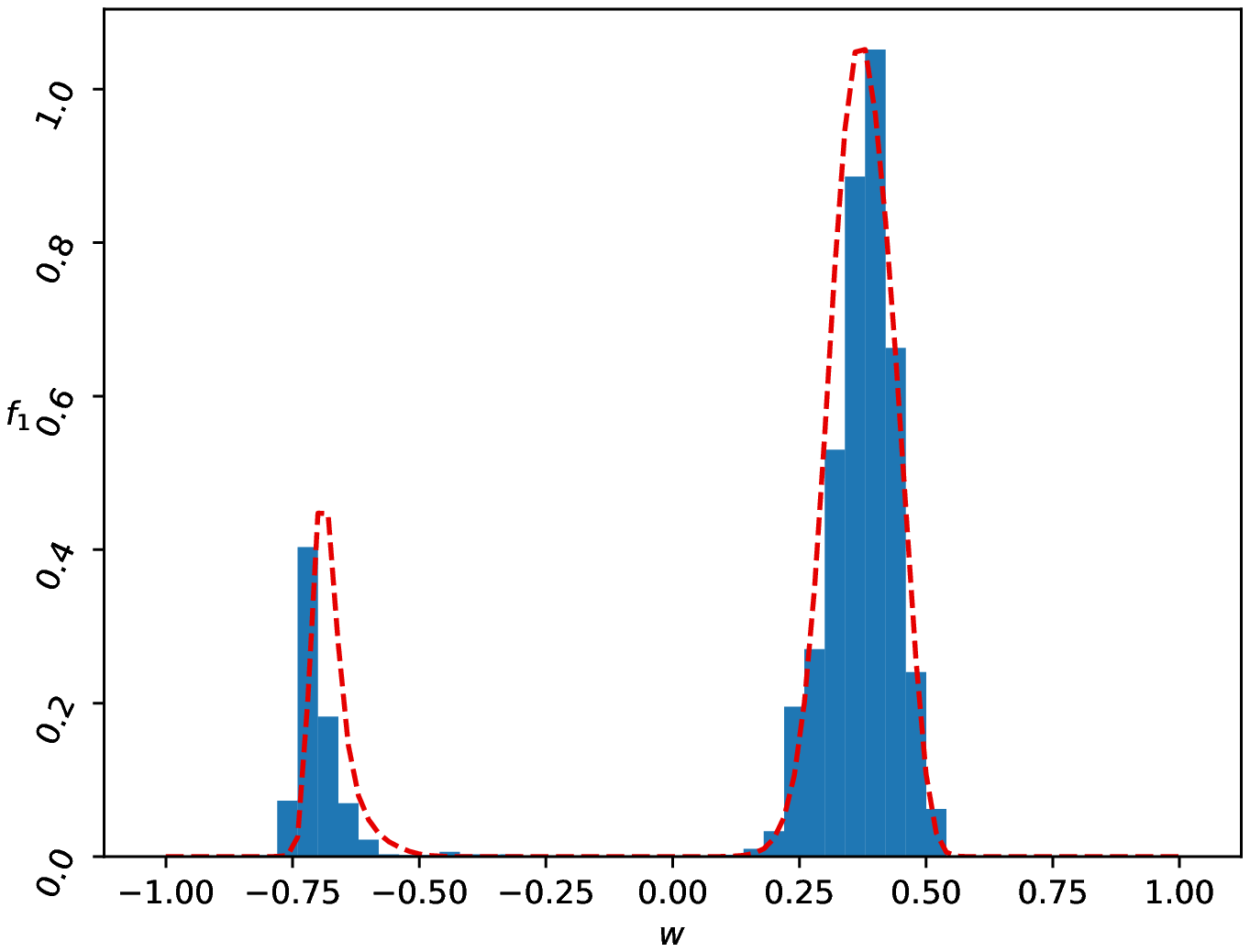}
            \caption{65+ year-old, remain}
	    \end{subfigure}
	    \caption{Numerical simulation of 2019 UK General Election in East of England: histogram representing the distribution of the individuals in the DSMC, with the results from a similar Fokker-Planck FEM solution superimposed as red dashed lines.}	
	    \label{fig:4.3analysis:MonteCarlo}
    \end{figure}
    
    The results of this comparative simulation are presented in Figure~\ref{fig:4.3analysis:MonteCarlo}. As we can observe, the results of the DSMC (the blue bars) and the FEM (the red dashed lines) coincide well. This suggests that using the FEM of the Fokker-Planck system is reasonable. We therefore present only the numerical experiments of the Fokker-Planck system in the remainder of this section.
    
    \subsubsection{Numerical Results: investigating the effect of data-driven $P_{char}$}
    
\begin{figure}[h]
	    \begin{subfigure}[b]{0.5\linewidth} 
		    \includegraphics[scale=0.4]{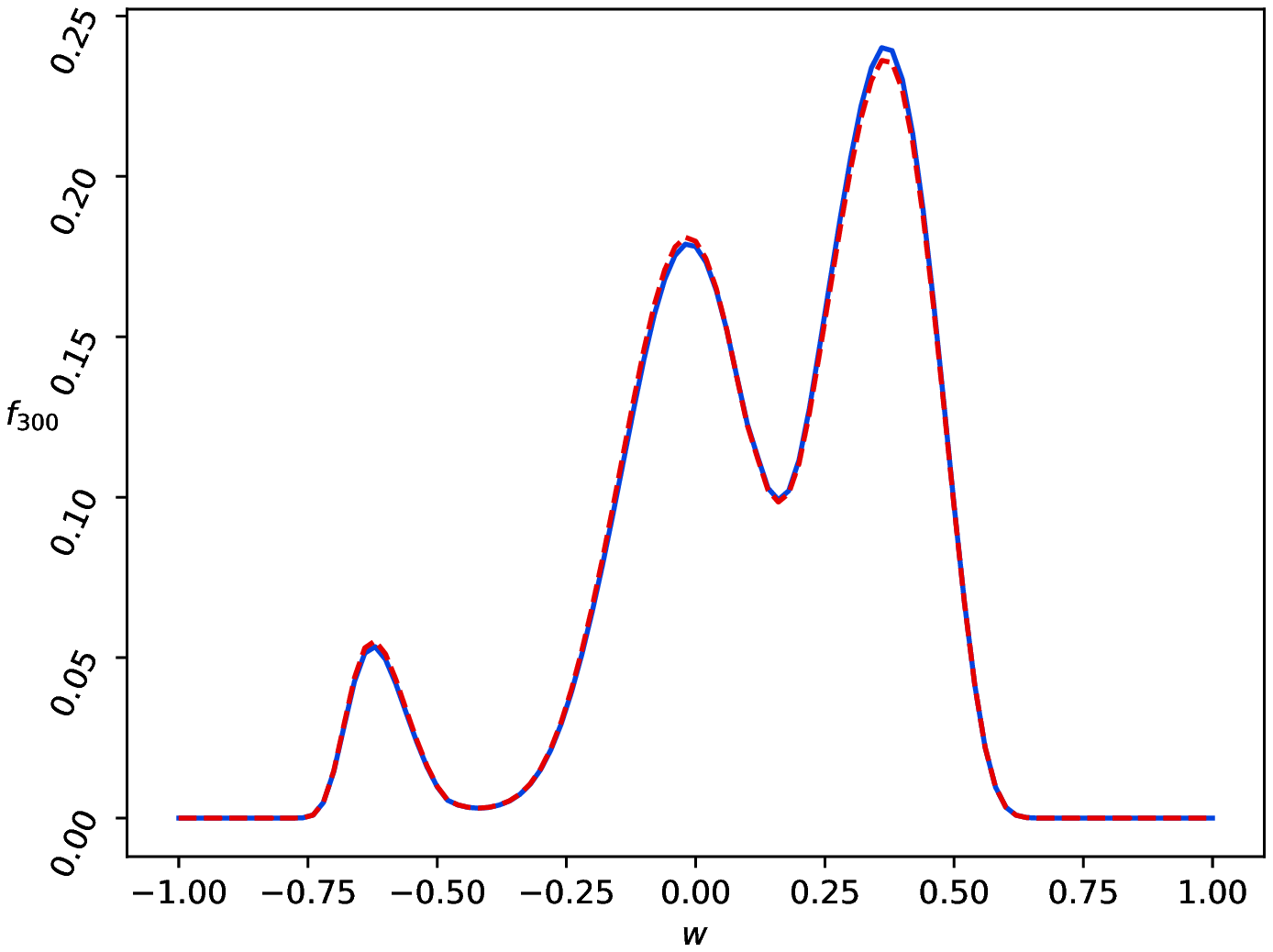}
    		\caption{Opinion Distribution with $P_{char}$ given by \eqref{eq:ourP}}
    		\label{fig:4.3analysis:comp:me}
	    \end{subfigure}
		    \hspace{0.04cm}
    	\begin{subfigure}[b]{0.5\linewidth} 
	    	\includegraphics[scale=0.4]{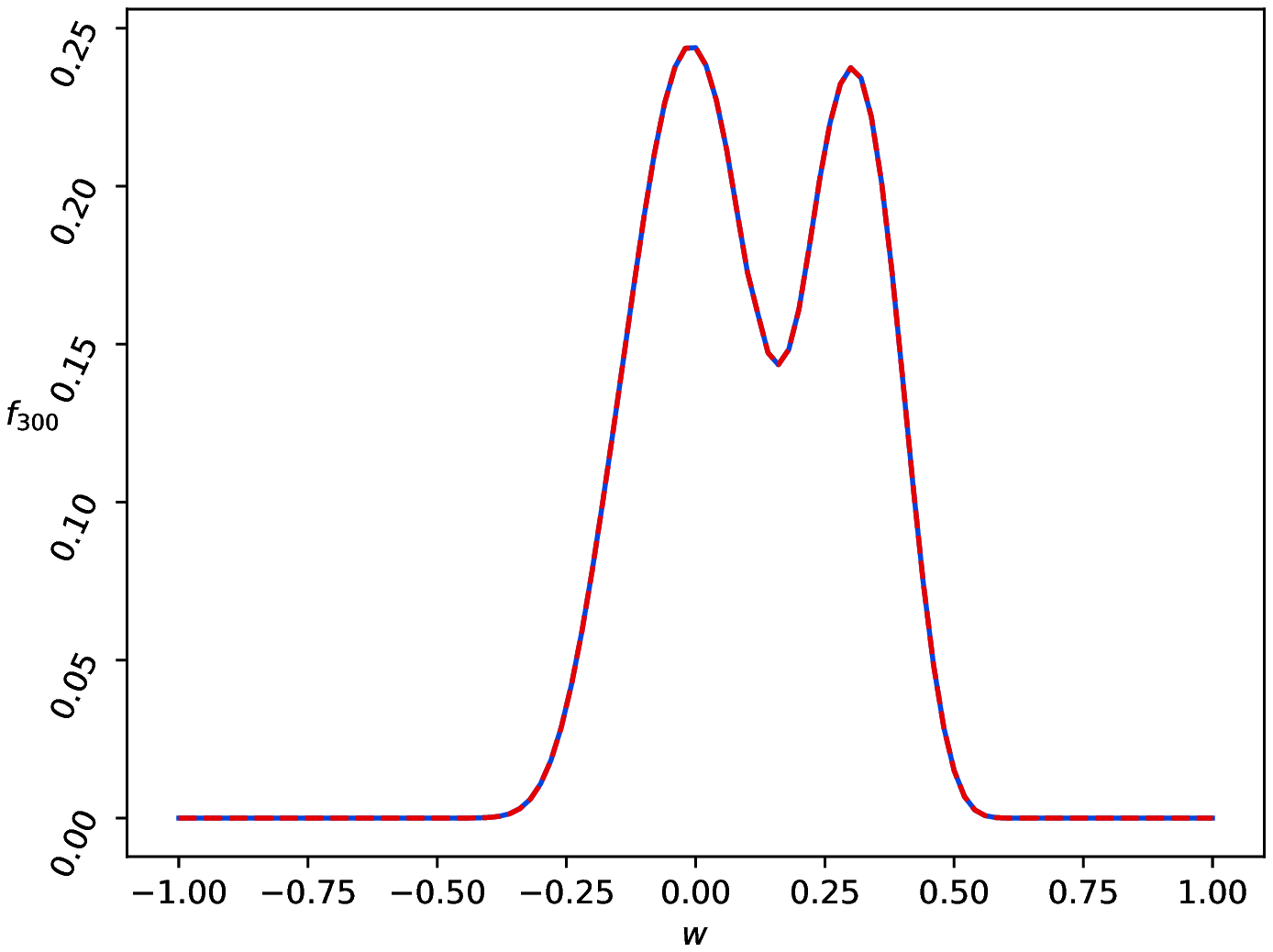}
		    \caption{Opinion Distribution with $P_{char} \equiv 1$ \phantom{lllllllllllllllllll}}
		    \label{fig:4.3analysis:comp:bertram}
	    \end{subfigure}
    	\caption{A comparison between the 18-24 year-old species in the East of England. \eqref{fig:4.3analysis:comp:me} are the results of the simulation with our choice of $P_{char}$, and \eqref{fig:4.3analysis:comp:bertram} are the results from the simulation with $P_{char}\equiv 1$.}
    	\label{fig:4.3analysis:comp}
    \end{figure}

    \noindent In our second numerical experiment, we illustrate the effect of our choice of $P_{char}$ which is chosen from demographic voter intention data (see Section \ref{sec:Deriv}.\ref{sec:Deriv:PFunction}) when compared to a model with $P_{char} \equiv 1$, similar to that presented by \textit{D\"uring et al.} \cite{bib:During:strongleaders}. To this end, we run various Fokker-Plank simulations with uniform initial density, with species' masses (see  Table~\ref{tab:masses} in Section~\ref{sec:Numerical}), looking at the long-time densities for $\tau_{iL} = 0.05$, $\tau_{ij} = 2.5$, $\lambda = 0.033$, $\Delta s = 0.1$ and $r = 0.85$. We run the model for 300 time steps to ensure the resulting distribution is the limiting distribution. We expect the formation of at least two peaks from our model, one within the influence of Labour ($[-1,-0.55)$) and one within the Conservative Influence ($[0.25,0.55)$). We also expect that, in our model, the species have distinct, yet similar, densities whereas in the model with $P_{char} \equiv 1$, we expect the densities to be identical. This is confirmed in our numerical simulations in Figure~\ref{fig:4.3analysis:comp}.
    
    
    In this simulation, we observe that a small peak appears in the influence of Labour in the case where our choice of $P_{char}$ is used, and the difference between the species' densities is very similar but there is a small, but distinct difference between Leave and Remain distributions, displayed as red and blue.

    
    From these simulations it is clear that the advection terms, through the $\mc{K}$ operators, are dominant in this type of model and so we see a net movement towards the large parties, from the compromise function $P_{char}$ and towards the centre of the domain, from the localisation function $P_{loc}$.  We believe that advection towards the centre of the two popular peaks at Labour ($-0.7$) and the Conservatives ($0.4$), that arises from the localisation function, leads to the success of the central parties. This then leads to the eclipse of the smaller parties, yielding only three peaks. This effect may be removed with the use of a smaller  interaction radius $r$ and rearranging of the parties.

 \label{sec:Numerical:Discussion}

  \subsection{Numerical results for 2019 UK general election's ``Red Wall''}

\begin{figure}[h]
	    \centering
	    \begin{subfigure}[b]{0.45\linewidth} 
	        \centering
		    \includegraphics[scale=0.65]{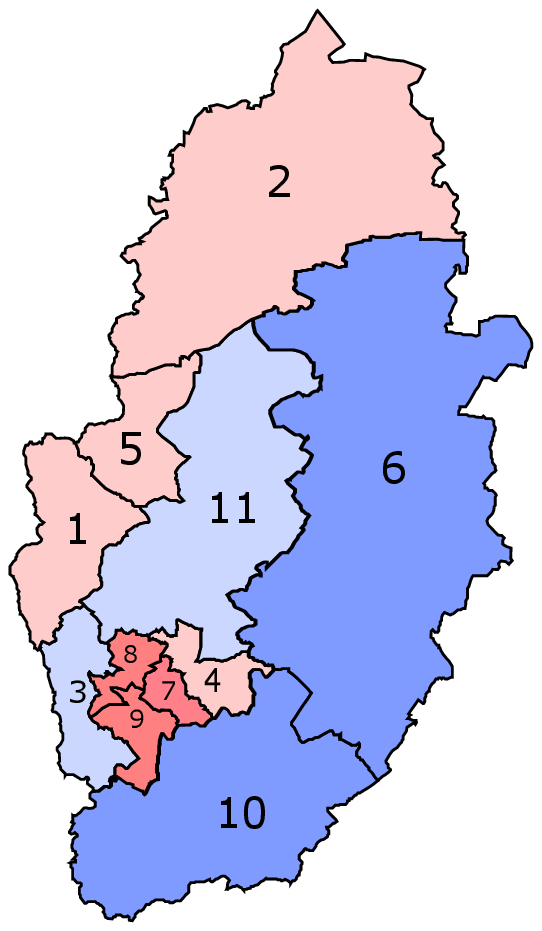}
    		\caption{Result of the 2015 General Election in Nottinghamshire}
    		\label{fig:4.3analysis:Not:2015act}
	    \end{subfigure}
    \hspace{0.04cm}
    	\begin{subfigure}[b]{0.45\linewidth} 
    	\centering
	    	\includegraphics[scale=0.65]{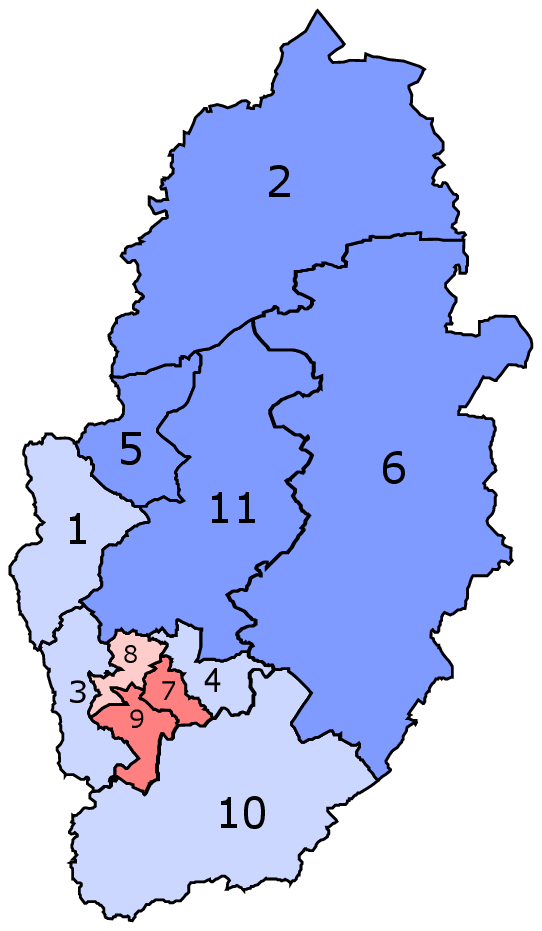}
		    \caption{Result of the 2019 General Election in Nottinghamshire}
		    \label{fig:4.3analysis:Not:2019act1}
	    \end{subfigure}
    	\caption{Results of the General Elections in 2015 \eqref{fig:4.3analysis:Not:2015act} and 2019 \eqref{fig:4.3analysis:Not:2019act1}. The colours represent the party to which the elected MP belongs with blue for Conservative and red for Labour. The light colours are where the total popular vote for the elected MP was below 50\% and the dark colours for greater than 50\%.}
    	\label{fig:4.3analysis:Not1}
    \end{figure}
\begin{figure}[h]
	    \centering
	    \begin{subfigure}[b]{0.45\linewidth} 
	        \centering
		    \includegraphics[scale=0.65]{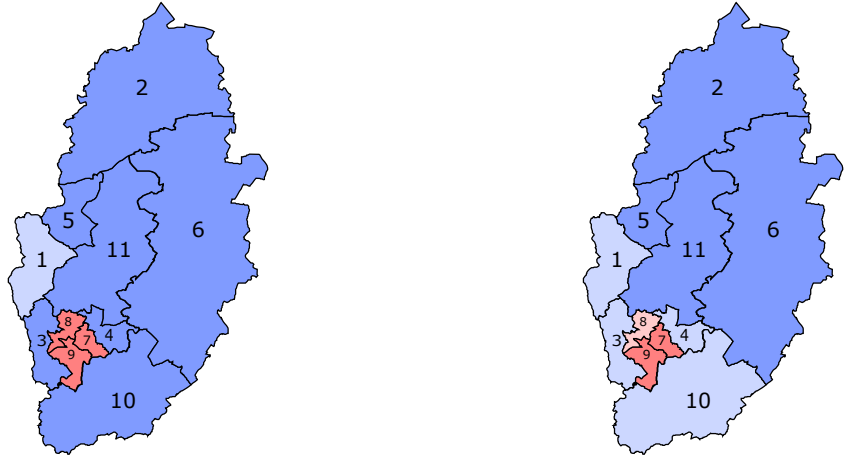}
    		\caption{Predicted Result of the 2019 General Election in Nottinghamshire}
    		\label{fig:4.3analysis:Not:2019predicted}
	    \end{subfigure}
    \hspace{0.04cm}
    	\begin{subfigure}[b]{0.45\linewidth} 
    	\centering
	    	\includegraphics[scale=0.65]{nottinghamshire2019actual.eps}
		    \caption{Recorded result of the 2019 General Election in Nottinghamshire}
		    \label{fig:4.3analysis:Not:2019act2}
	    \end{subfigure}
    	\caption{Numerical simulation results predicted by our proposed kinetic model, compared to the 2019 General Election result in Nottinghamshire. The colours represent the party to which the elected MP belongs with blue for Conservative and red for Labour. The light colours are where the total popular vote for the elected MP was below 50\% and the dark colours for greater than 50\%.}
    	   \label{fig:4.3analysis:Not:Results2}
    \end{figure}

    Finally, we apply this model to a real-world situation, to show a proof of concept that models of this type can be used with sufficient data. 
    As discussed in Section~\ref{sec:Numerical}.\ref{sec:Numerical:DataUsage}, we run Fokker-Planck simulations for the county of Nottinghamshire using the parameter values from the simulations in the East of England simulations above namely: $\tau_{iL} = 0.05$, $\tau_{ij} = 2.5$, $\lambda = 0.033$, $\Delta s = 0.1$ and $r = 0.85$, where we use the same method presented in Section~\ref{sec:Numerical}.\ref{sec:Numerical:DataUsage}(\ref{sec:Data:EOECaseStudy}), to generate our initial masses, namely by taking the product of the relevant columns in our data tables (an example of which is Table \ref{tab:PFunction:examplechar}) element-by-element for both age and self-reported 2016 Brexit vote, then dividing this mass evenly between the populations obtained from the 2011 Census, and to regionalise the voter intention data, by taking the columns of the national voter intention data and the results from the 2017 General Election and multiplying element-by-element. We also generate the initial densities in the same way as that section, that is we use sums of characteristic functions.
         The Nottinghamshire constituencies in Figures \ref{fig:4.3analysis:Not1} and \ref{fig:4.3analysis:Not:Results2} are labeled by numbers: (1) Ashfield, (2) Bassetlaw, (3) Broxtowe, (4) Gedling, (5) Mansfield, (6) Newark, (7), (8) and (9) Nottingham East, North and South respectively, (10) Ruscliffe and (11) Sherwood. Colour-shading indicates the outcome of the 2015 and 2019 General Election in each constituency.
        
  We see the shift of the northernmost three constituencies, Ashfield, Bassetlaw and Mansfield, from Labour, as a part of the ``Red Wall", to Conservative in Figure \ref{fig:4.3analysis:Not1} in the 2019 General Elections. The so-called ``fall of the Red Wall" was seen as a decisive factor in the 2019 victory for the Conservatives \cite{bib:Guardian, bib:BBC} since many Labour MPs were elected in these constituencies, forming the backbone of the Labour party throughout most of the $20^{th}$ century.

    This is a good example for testing our new model since, from the recorded results of the 2015 and 2019 General elections \cite{bib:HOC:2017GE, bib:HOC:2019GE}, there are a few constituencies which elected a Labour MP in 2015 and a Conservative MP in 2019 (Ashfield, Bassetlaw, Mansfield and Gedling), a few constituencies that elected a Conservative MP in both elections (Broxtowe, Newark, Rushcliffe and Sherwood), and some that elected a Labour MP in both elections (Nottingham East, North and South).
    
    As can be seen in the numerical simulation results in Figure~\ref{fig:4.3analysis:Not:2019predicted}, our model correctly assigns a ``winner", the party with the highest influence mass (aggregated over all eight demographic species), in each constituency. The results from the kinetic model predict a larger winning majority than the actual results in some constituencies, however, these quantitative differences are less relevant in these first-past-the-post elections, demonstrating the utility of this model. We believe that further refinement of this model along with more current data may alleviate this issue, which we leave to future research.

\begin{figure}
	    \begin{subfigure}[b]{0.5\linewidth}
		    \includegraphics[scale=0.4]{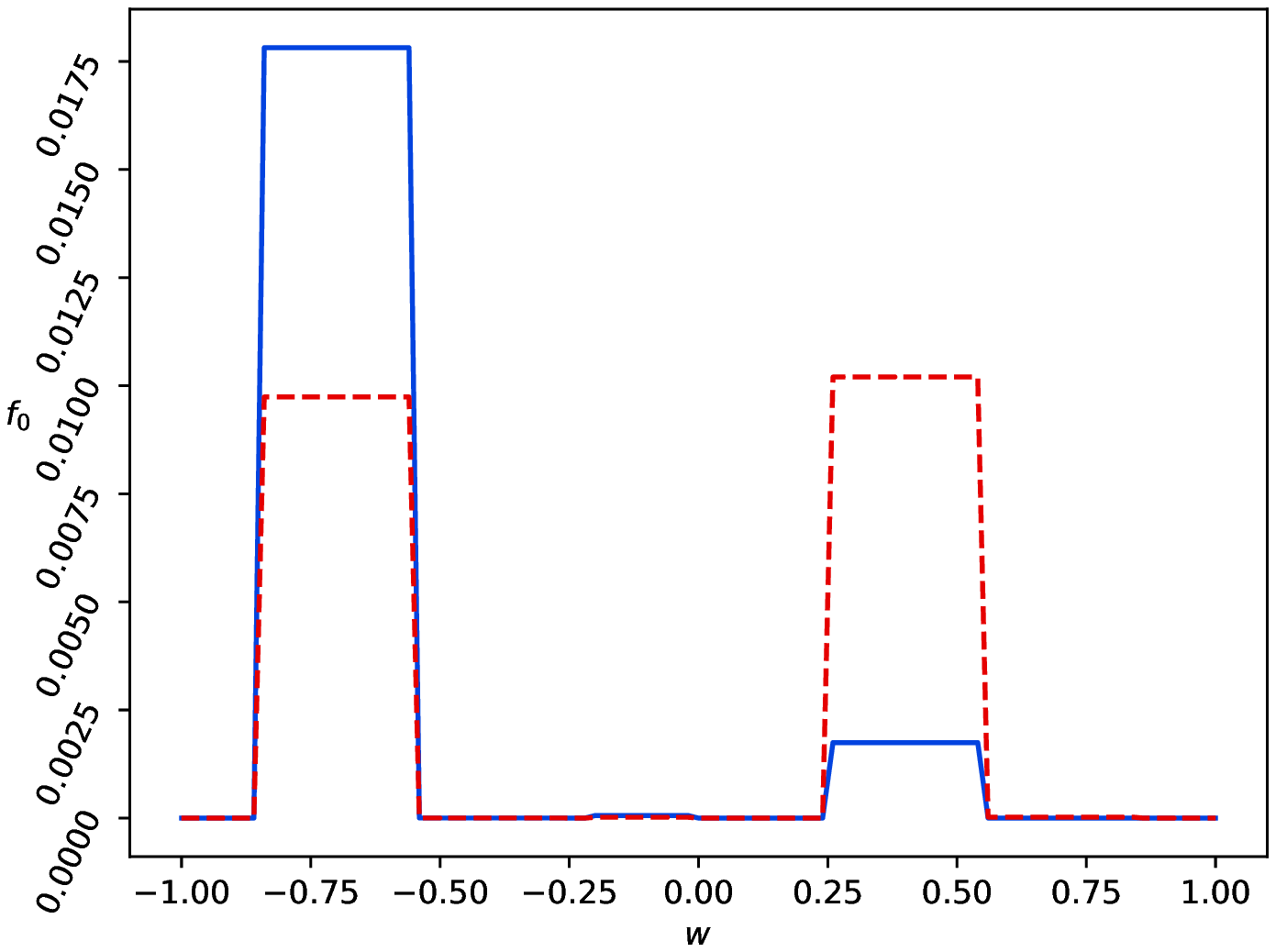}
		    \caption{18--24 year-old}
	    \end{subfigure}
		\hspace{0.04cm}
	    \begin{subfigure}[b]{0.5\linewidth}
		    \includegraphics[scale=0.4]{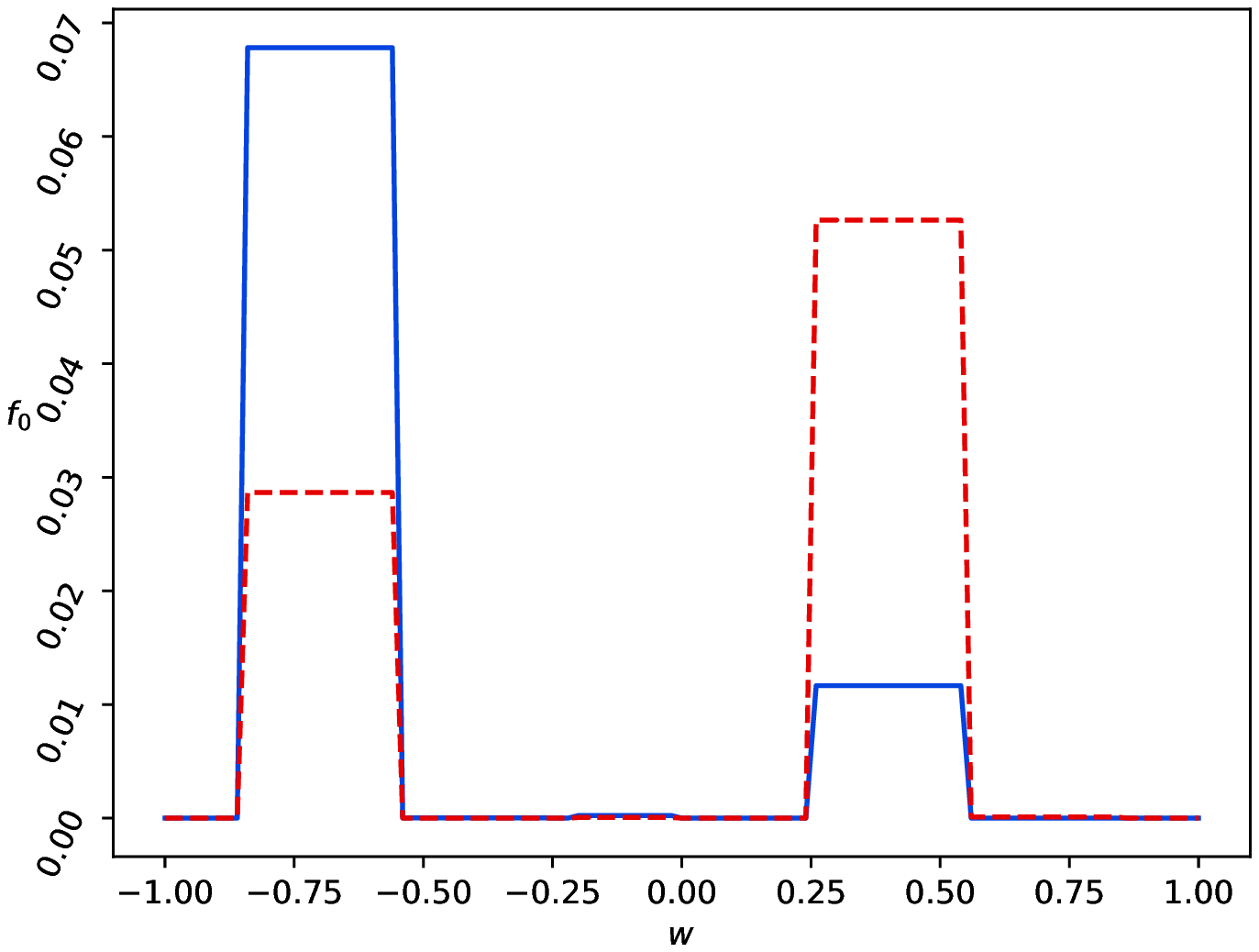}
		    \caption{25--49 year-old}
	    \end{subfigure} 
    
    	\begin{subfigure}[b]{0.5\linewidth}
	    	\includegraphics[scale=0.4]{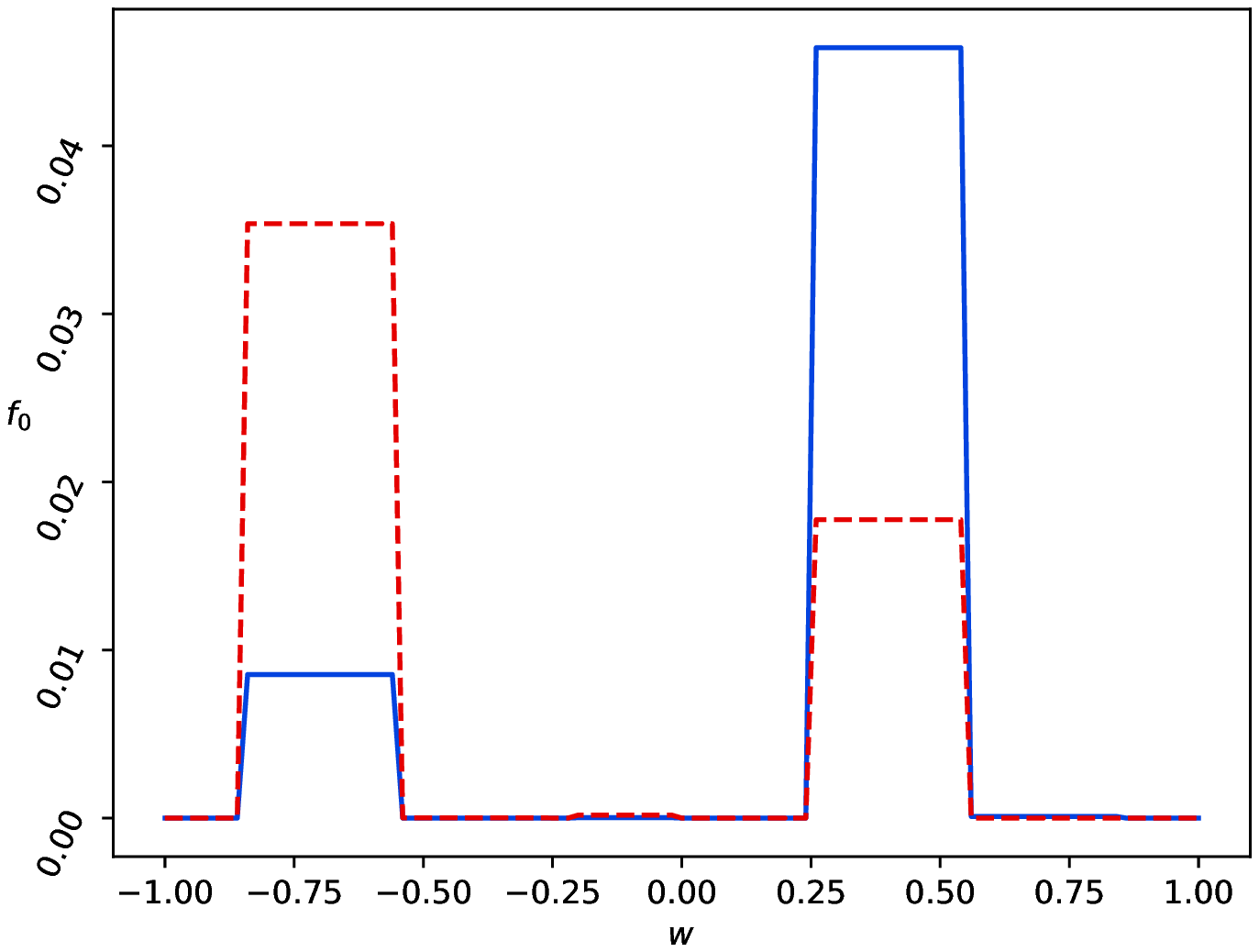}
		    \caption{50--64 year-old}
	    \end{subfigure} 
	    \hspace{0.04cm}
	    \begin{subfigure}[b]{0.5\linewidth}
		    \includegraphics[scale=0.4]{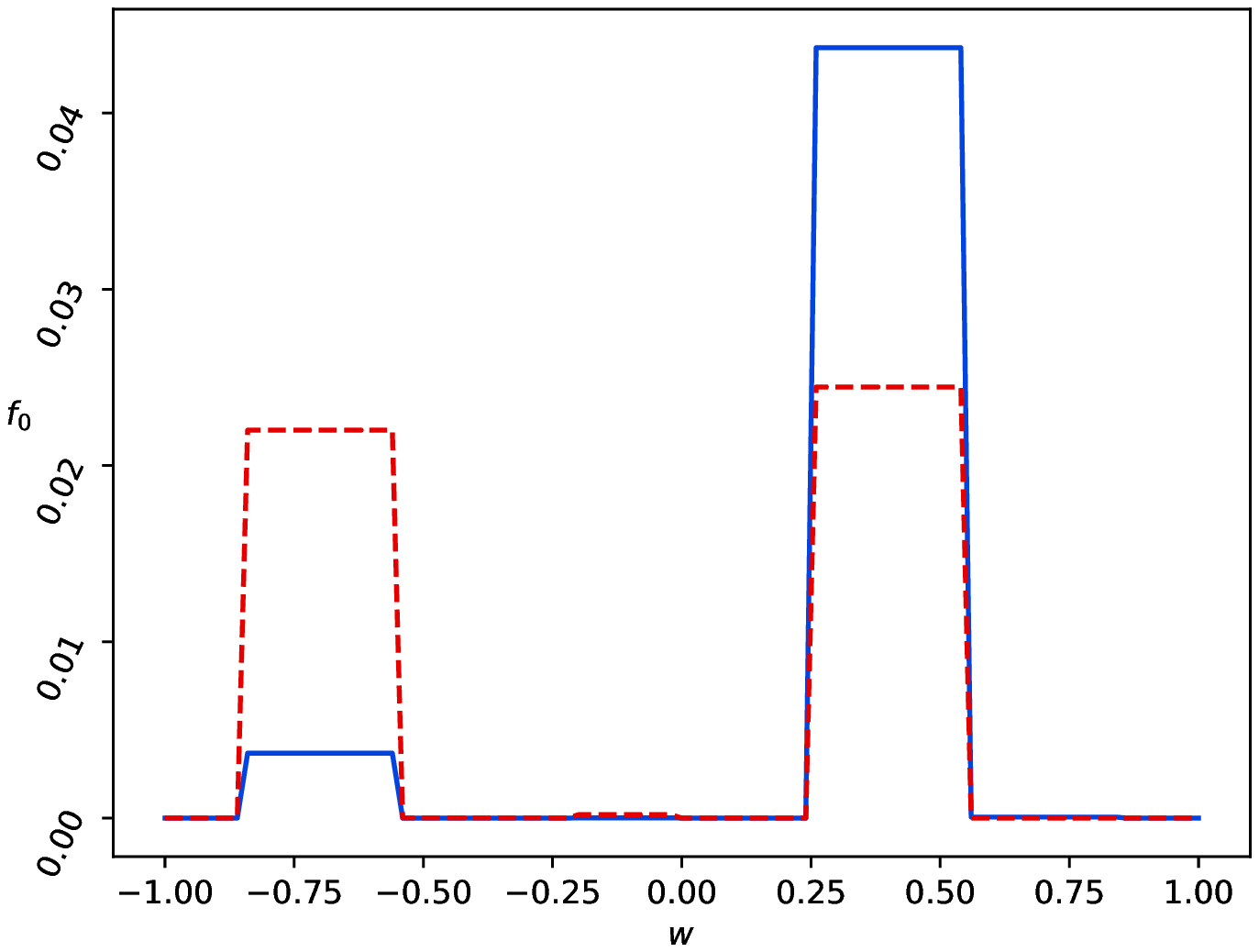}
            \caption{65+ year-old}
	    \end{subfigure}
	    \caption{Initial densities for the eight follower species in the constituency of Bassetlaw from the 2015 General Election Results. Each graph shows the initial density for the leave (solid blue) and the remain (dashed red) characteristic of an age group.}	
	    \label{fig:4.3analysis:demographicbassics}
    \end{figure}

\begin{figure}
	    \begin{subfigure}[b]{0.5\linewidth}
		    \includegraphics[scale=0.4]{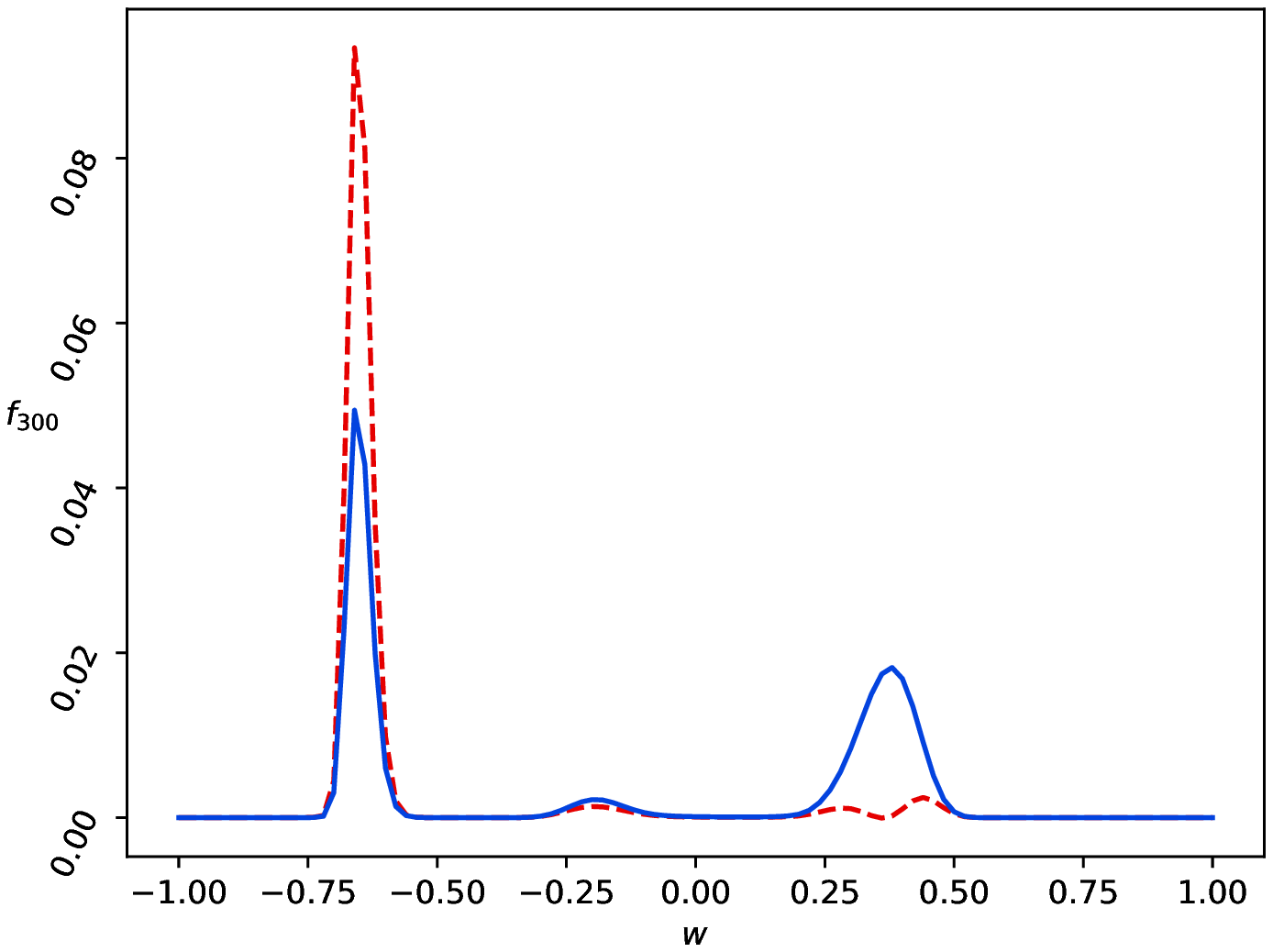}
		    \caption{18--24 year-old}
	    \end{subfigure}
		\hspace{0.04cm}
	    \begin{subfigure}[b]{0.5\linewidth}
		    \includegraphics[scale=0.4]{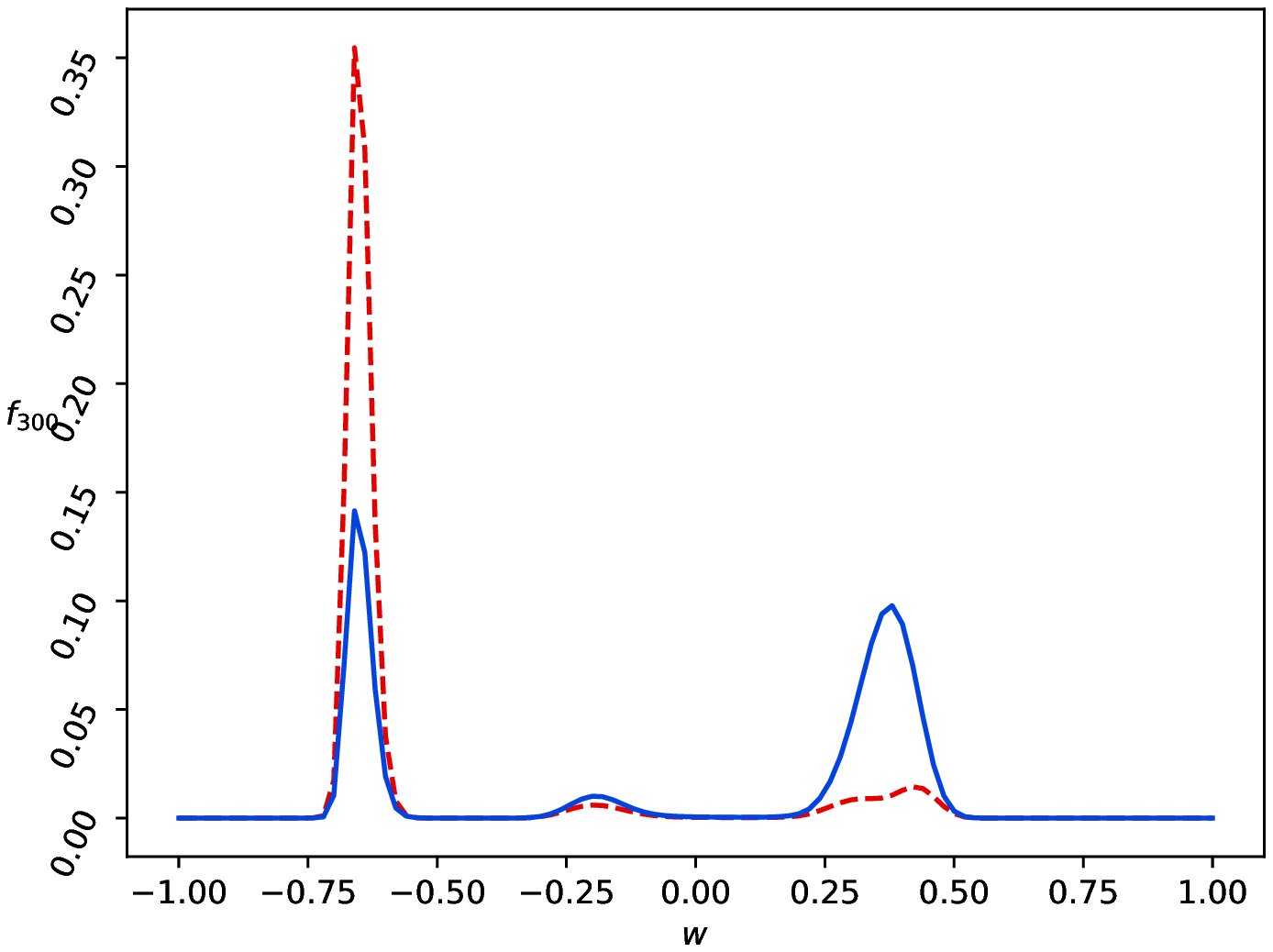}
		    \caption{25--49 year-old}
	    \end{subfigure} 
    
    	\begin{subfigure}[b]{0.5\linewidth}
	    	\includegraphics[scale=0.4]{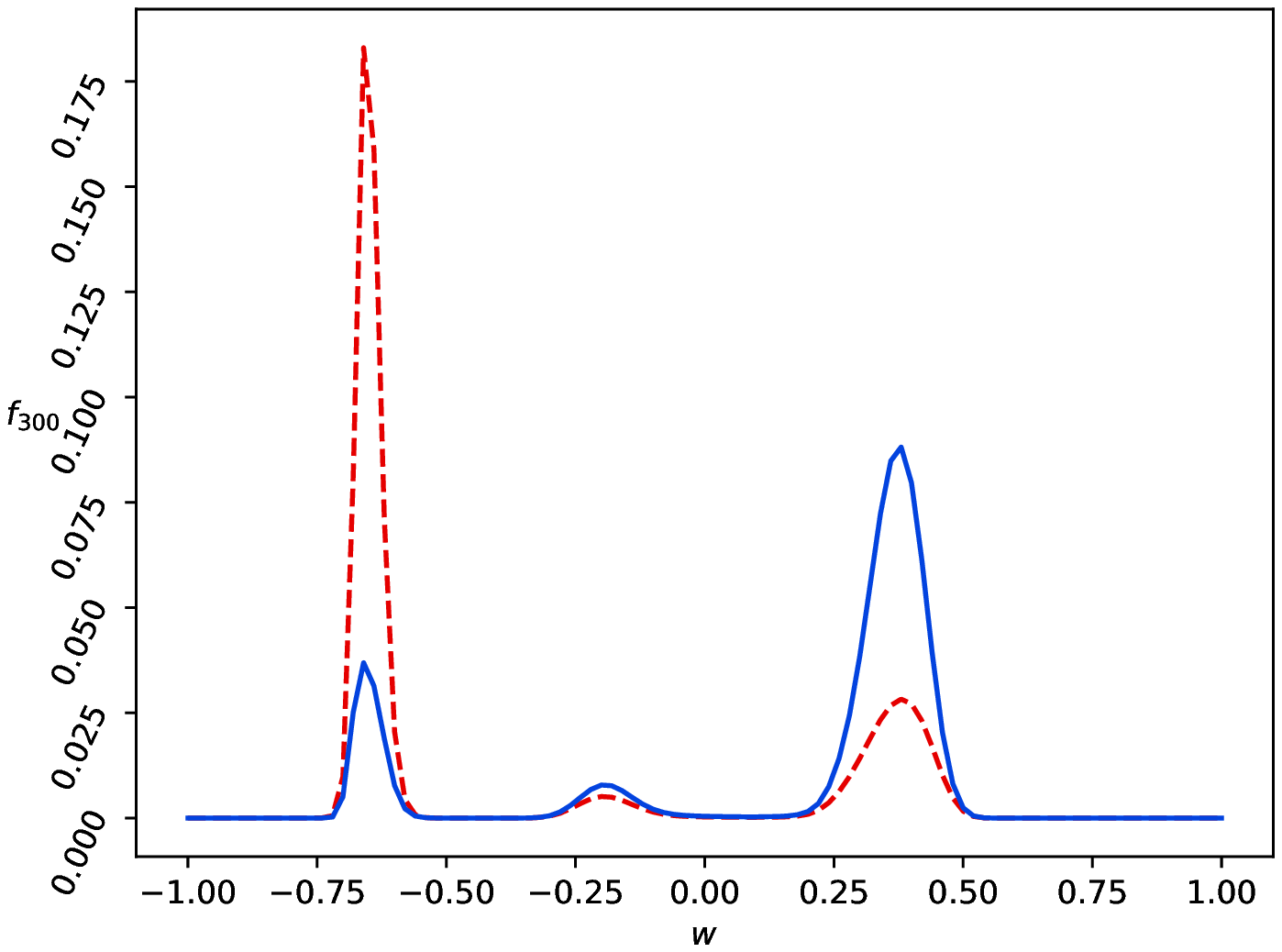}
		    \caption{50--64 year-old}
	    \end{subfigure} 
	    \hspace{0.04cm}
	    \begin{subfigure}[b]{0.5\linewidth}
		    \includegraphics[scale=0.4]{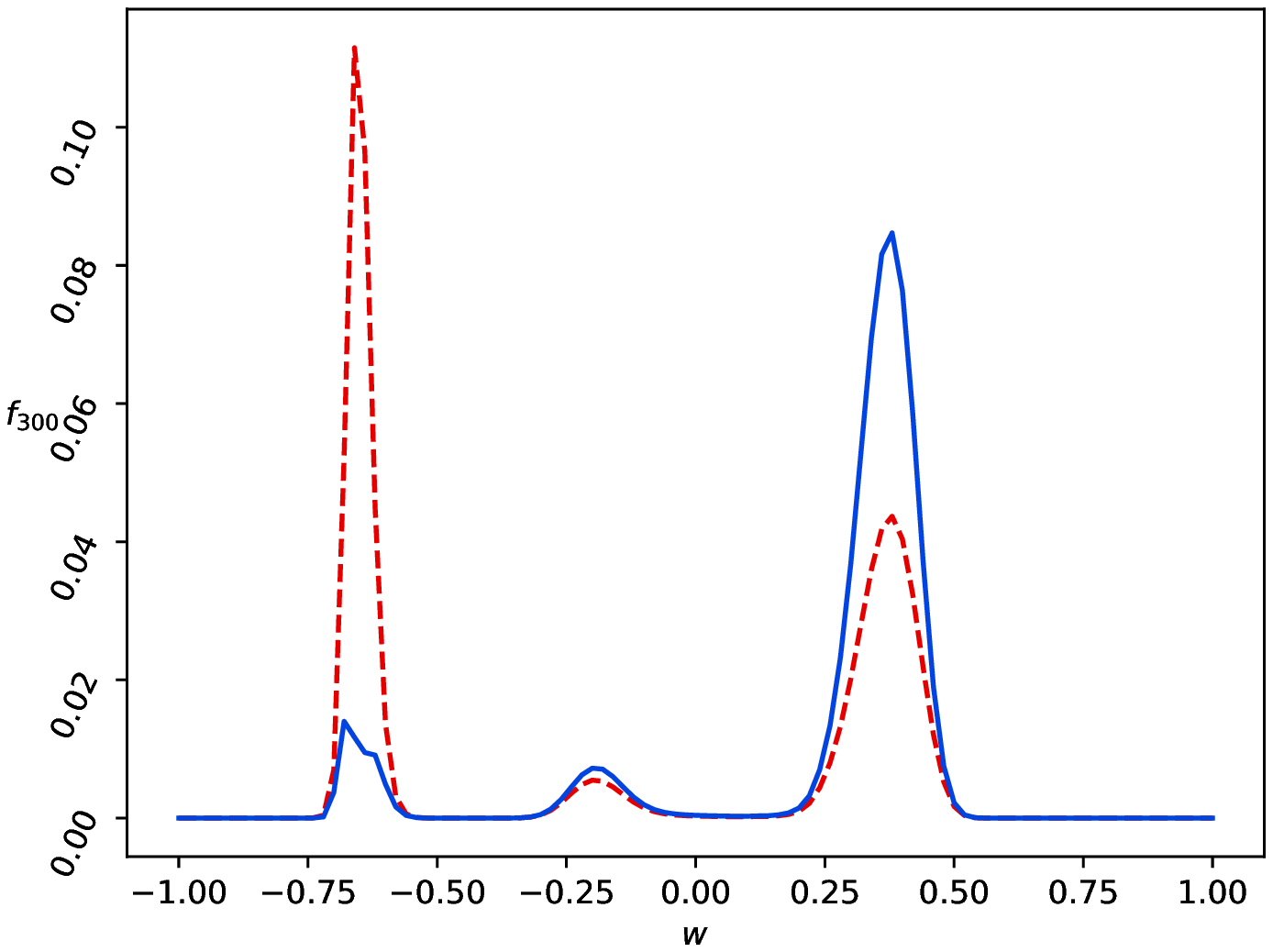}
            \caption{65+ year-old}
	    \end{subfigure}
	    \caption{Predicted Results from the model for the constituency of Bassetlaw. Each graph shows the predicted density for the leave (solid blue) and the remain (dashed red) characteristic of an age group.}	
	    \label{fig:4.3analysis:demographicbass}
    \end{figure}

    We now look more in depth at the constituency of Bassetlaw, for which the initial eight follower densities (generated from the 2015 General Election Results) are presented in Figure~\ref{fig:4.3analysis:demographicbassics}, and the predicted results from our model are presented in Figure~\ref{fig:4.3analysis:demographicbass}. As we can see, the distinction between the different demographics is quite clear, the younger demographics -- those with ages 18-24 years-old and 25-49 years-old  -- show the largest difference in initial to predicted opinion, whereas the two older demographics show little change from the initial density. Since the demographics with the largest total masses are the 25-49 year-olds (both leave and remain), this shift mass from the Labour influence interval to the Conservative influence interval may have been the cause for the entire constituency to vote for a Conservative MP in the 2019 General Election.

    We additionally note that the older demographics masses are either split somewhat evenly over the Labour and Conservative influence intervals (in the case of Remain voters) or the majority of the mass in the Conservative influence interval (in the case of Leave voters) as compared to the higher mass in the Labour influence interval in general for the younger demographics. Figure~\ref{fig:4.3analysis:demographicbass} shows a clear correlation between age and Conservative voting and self-reported Brexit vote and Conservative voting predicted in our model -- a correlation that has been observed since the election \cite{bib:YouGov:howbritainvoted}.

 \label{sec:Numerical:RedWall}

\newpage

\section{Conclusion}

    We have proposed a kinetic model for opinion formation which takes into account the effect of voter demographics and other socio-economic factors on the opinion formation dynamics. Using methods of kinetic theory, from microscopic opinion consensus dynamics, we have derived a mesoscopic Boltzmann-type system and, in a quasi-invariant limit, a macroscopic Fokker-Planck system. In this model, the followers' compromise process is biased towards their preferred opinions, influenced by their demographic factors like age and their behaviour in previous election or referenda, as suggested by the results of real-world voter polls and post-election analyses. The use of voter intention data in the compromise function yields  interpretable model parameter functions, unlike in alternative approaches, as for example a process of calibrating mathematical model parameters to obtain (or be close to) a desired target distribution.
    Application of this model to the 2019 UK general election, using voter intention data and previous election data illustrates the potential of such models and also highlights the importance of quality data. For Nottinghamshire, part of the so-called ``Red Wall'', where there was a strong shift in traditional Labour strongholds to the Conservative party, the model has been able to correctly assign all parliament seat winners. This election outcome is explained among other factors by the strong support for ``Leave'' in the 2016 Brexit referendum vote in these constituencies and the positioning of the two large parties in the EU membership question -- an effect which our model is able to capture and include in the opinion formation dynamics.
    \label{sec:Conclusion}


\section*{Data Access}
    YouGov voter intention data is available at \cite{bib:YouGov:voterintention}.
    2011 Census data is available at \cite{bib:ONS:2011census}.
    Results of the 2015, 2017, 2019 UK general elections are available at \cite{bib:HOC:2015GE,bib:HOC:2017GE,bib:HOC:2019GE}.
    Results of the 2016 EU referendum are available at \cite{bib:ElectoralComission:2016Brexit}.
    The code used to run the simulations in Section~\ref{sec:Numerical} can be found at: https://github.com/the-ollie-wright/kofm-pep.





\printbibliography

\end{document}